\documentclass[aps,prl,reprint,superscriptaddress, longbibliography, nofootinbib]{revtex4-2}

\usepackage{blindtext}
\usepackage{graphicx}
\usepackage{amsfonts}
\usepackage{amsmath}
\usepackage{bbold}
\usepackage{physics}
\usepackage[caption=false]{subfig}
\usepackage{float} 
\usepackage{xcolor}
\usepackage[export]{adjustbox}

\definecolor{dodgerblue}{HTML}{1E90FF}
\definecolor{lightdodgerblue}{HTML}{4dbff7}

\usepackage[colorlinks=true,
    linkcolor=blue,
    citecolor=red,
    filecolor=magenta,      
    urlcolor=cyan,
    pdftitle={GoldenRuleCorrection},]{hyperref}

\usepackage{comment}
\usepackage[normalem]{ulem}  

\makeatother

\begin{document}

\title{Breakdown of Fermi's Golden Rule in 1d systems at non-zero temperature}

\author{Thomas Young} 
  \affiliation{Department of Physics, King’s College London, Strand WC2R 2LS, UK}

\author{Jerome Lloyd}
\affiliation{Department of Theoretical Physics, University of Geneva, Geneva, Switzerland}

\author{Curt von Keyserlingk} 
 \affiliation{Department of Physics, King’s College London, Strand WC2R 2LS, UK}

\date{\today}
\begin{abstract}
In interacting quantum systems, the single-particle Green's function is expected to decay in time due to the interaction induced decoherence of quasiparticles. In the limit of weak interaction strengths ($\Delta$), a naive application of Fermi's Golden Rule (FGR) predicts an  $\mathcal{O}(\Delta^{2})$  quasiparticle decay rate. However, for 1d fermions on the lattice at $T>0$, this calculation gives a divergent result and the scaling of the quasiparticle lifetime with interaction strength remains an open question. In this work we propose a solution to this question: combining numerical simulations using the recently introduced dissipation-assisted operator evolution (DAOE) method, with non-perturbative diagrammatic re-summations, we predict a logarithmic enhancement of the quasiparticle decay rate $\tau^{-1} \sim \Delta^{2} \log \Delta^{-2}$. We argue that this effect is present in a wide variety of well-known weakly interacting quantum fermionic and bosonic systems, and even in some classical systems, provided the non-interacting limit has quasiparticles with a generic dispersion. 
\end{abstract}

\maketitle

\paragraph{Introduction.---}

A long-standing aim of condensed matter theory is to predict and understand the behavior of correlation functions in non-equilibrium many-body systems.  The single-particle Green's function, which quantifies the coherence of quasiparticles, is perhaps the simplest such quantity. It is expected to decay rapidly at non-zero temperatures in interacting systems, as quasiparticles are rapidly dephased through interactions with a many-body background \cite{Sachdev_2023,mcculloch_2025}. 

The present work aims to understand how the quasiparticle lifetime scales with interaction strength ($\Delta$) in the limit of weak interactions, and at non-zero temperatures. While a cursory application of Fermi's Golden Rule (FGR) suggests a decay rate $\tau^{-1}\sim \Delta^2$, a more careful calculation shows that the rate diverges at leading order in perturbation theory, in one spatial dimension. A finite answer is expected to emerge when all orders in pertubation theory are included. In this Letter we argue that the resulting rate acquires a logarithmic enhancement 
\begin{equation}
\tau^{-1} \sim \Delta^2\log\Delta^{-2}.\label{eq:maintauinv}
\end{equation}
This effect is general in weakly-interacting 1d systems, appearing in fermionic lattice models with generic dispersions, most continuum fermion models\footnote{Curiously, due to a Pauli exclusion effect, the enhancement is absent in a previously studied model of spinless fermions in the continuum \cite{bertini_essler}.}, as well as lattice and continuum bosonic models and their classical limits. For concreteness, we focus most of our discussion on the canonical model of spinless fermions on the 1d lattice (Eq.~\eqref{eqn:Hamiltonian}), where the only requirement for the logarithmic enhancement is for the energy density and filling to be nonzero ($T>0, |\beta\mu|<\infty$). We also confirm the scaling in a non-integrable 1d fermion lattice model, and a classical model with `bosonic' fields.

Quasiparticle lifetimes have of course been studied extensively in prior literature, prominently in discussions of Fermi liquids, and other low-temperature models \cite{Sachdev_2023}. On the other hand, the weakly interacting 1d $T>0$ setting is, somewhat surprisingly, an understudied regime, which may account for the logarithmic enhancement having been overlooked. Hints of this effect have been seen in low-temperature calculations \cite{Glazman_2007,Glazman_2012,Pereira_2009}, and in an investigation of Coulomb drag in a 1d ladder \cite{Coulomb_drag_2019}, but so far have not been studied systematically.

The single-particle Greens function plays a prominent role in connection to experiment, appearing in particular in spectroscopic measures of condensed matter systems \cite{berthod2018spectroscopic}. Additionally, in systems where transport is governed by quasiparticles, it is natural to expect that the enhancement Eq.~\eqref{eq:maintauinv} shows up in measurable transport properties. We will return to this point in the discussion. Our work thus provides a new qualitative prediction for quasiparticle lifetimes, and possibly for transport, in a wide range of experimentally relevant quantum systems. Confirming the logarithmic enhancement Eq.~\eqref{eq:maintauinv} is a worthy challenge for near-term quantum simulation experiments, as well as for the numerical algorithms being developed to simulate quantum dynamics \cite{Bardarson2022,Bardarson2024,White2023,Banuls2025,DAOE1,DAOE2}. 

\paragraph{The model.---}
Consider lattice fermions in 1d 
\begin{equation}
    H = -\frac{1}{2}\sum_{i} \left(f_{i+1}^{\dagger}f^{\null}_{i} + f_{i}^{\dagger}f^{\null}_{i+1}\right) + \Delta \sum_{i} n_{i+1}n_{i},
\label{eqn:Hamiltonian}
\end{equation}
where $f_{i}$ is the fermion annihilation operator for site $i$ and $n_{i} = f_{i}^{\dagger}f^{\null}_{i}$. We focus on the limit of weak interaction strength ($\Delta$). This model is integrable, but we choose to work with it for its conceptual simplicity. We confirm most of our analytical and numerical results for Eq.~\eqref{eqn:Hamiltonian} with added integrability-breaking terms, which suggests that the integrability of the model is unimportant for the logarithmic enhancement of scattering rate $\tau^{-1}$. 

We focus on the single-particle Green's function
\begin{equation}
\label{eqn:single-particle Green's function}
    G_{k}(t) = \langle f^{\null}_{k}(t)f_{k}^{\dagger}(0)\rangle,
\end{equation}
where $\langle \cdot \rangle$ denotes the thermal expectation value,  and $f_{k}(t) = e^{\mathrm{i}Ht}f_{k}(0)e^{-\mathrm{i}Ht}$ is a Heisenberg evolved operator. For simplicity we work at infinite temperature, but we argue that the physical effect under study persists provided $T>0,|\beta \mu|<\infty$ \cite{SM}. It is convenient to work with the Laplace transform of Eq.~\eqref{eqn:single-particle Green's function}
\begin{equation}
\label{eqn:Dyson equation}
    G_{k}(z) = \int_{0}^{\infty} dt \; e^{\mathrm{i} zt} G_{k}(t) \propto \frac{1}{z - \epsilon_{k} - \Sigma_{k}(z)},
\end{equation}
where $\epsilon_{k} = -\cos k$ and $\Sigma_{k}$ is the single-particle self-energy.

The temporal decay of the single-particle Green's function is encoded in the analytical structure of $G_{k}(z)$. For example, a pole at $z = \omega - \mathrm{i}\Gamma$ corresponds to a real time exponential decay $G_{k}(t) \sim e^{-\mathrm{i}\omega t - \Gamma t}$ with $\Gamma$ being identified as the single-particle decay rate. This picture is overly simplistic because in the 1d systems of interest, the single-particle Green's function need not decay exponentially (although it always appears to decay superpolynomially) \cite{mcculloch_2025}. Nevertheless we expect the solutions to $ z_{*} = \epsilon_{k} + \Sigma_{k}(z_{*})$, and in particular the imaginary part of $\Sigma_{k}(z_{*})$, to set the (inverse) characteristic timescale for the decay of Eq.~\eqref{eqn:single-particle Green's function}. Using the fact that the self-energy $ \Sigma_{k}(z)$ disappears at $\Delta=0$, we expect the on-shell self-energy
\begin{equation}
\label{eqn:Quasiparticle decay rate}
    \tau_{k}^{-1} =  -\;\text{Im}\; \Sigma_{k}(z = \epsilon_{k} + \mathrm{i}0^{+})
\end{equation}
to set the quasiparticle decay rate \cite{Sachdev_2023, Altland_Simons_2010}.  In general the self-energy is not known exactly but is approximated by truncating the diagrammatic series at low order in $\Delta$ (e.g., FGR) or by re-summing a subset of diagrams to all orders \cite{Buchhold2015,Rosenhaus_2019}. To begin, we demonstrate that the FGR approximation gives a divergent result for Eq.~\eqref{eqn:Hamiltonian}. 

The leading order contribution to the quasiparticle decay rate is second order in $\Delta$ (Eq.~\ref{eq:self-energy-leading}), hence FGR predicts
\begin{equation}
\label{eqn:Quasiparticle lifetime lowest order}
    \tau_{k}^{-1} \! \approx  \!-2\Delta^{2}\! \Im \! \int \frac{\mathrm{d}p\mathrm{d}q}{(2\pi)^{2}} \; \frac{(\cos q - \cos p)^{2}A_{k}(q,p)}{\epsilon_{k} - \epsilon_{k+q} - \epsilon_{k+p} + \epsilon_{k+q+p} + \mathrm{i}0^{+}}.
\end{equation}
where $A_{k}(q,p)= n_{p+q+k}(1-n_{q+k}-n_{p+k})+n_{q+k} n_{p+k}$ and the $n$ factors denote single-particle Fermi-Dirac distributions. 

The momentum integral diverges since the function appearing in the denominator, $\phi_{k}(q,p) = \epsilon_{k} - \epsilon_{k+q} - \epsilon_{k+p} + \epsilon_{k+q+p} $, has a quadratic pole, i.e., the equations
\begin{equation}
\label{eqn:Stationary phase conditions on shell}
    \phi_{k}(q,p) = \partial_{q}\phi_{k}(q,p) = \partial_{p}\phi_{k}(q,p) = 0
\end{equation}
are satisfied at two special points $(q_{*},p_{*}) = (0,\pi-2k),(\pi-2k,0)$. These solutions correspond to the incoming particle scattering into a particle-particle-hole triplet, where the outgoing particles/holes all move with the same group velocity \cite{Coulomb_drag_2019}. The quadratic pole is present for all $k$ except $k = \pm \pi/2$ where the divergence is nullified by the disappearance of the numerator in Eq.~\eqref{eqn:Quasiparticle lifetime lowest order}, a manifestation of Pauli-exclusion.  The logarithmic divergence in this diagram has been observed previously in the study of Coulomb drag \cite{Coulomb_drag_2019} in a specific setup of fermions on a ladder. It has also been observed in a calculation of quasiparticle lifetimes in spinless fermions at $T=0$ \cite{Pereira_2009}, for specific values of $k$. However, at zero temperature the divergence is ameliorated by a vanishing density of states when treated non-perturbatively within the Luttinger liquid paradigm \cite{Glazman_2007,Glazman_2012}. 

It is worth commenting that these divergences are rather generic.  Firstly, the diagram Eq.~\eqref{eq:self-energy-leading} is similarly divergent at any non-zero temperature and finite chemical potential; roughly, all that changes are that the divergent points are accompanied by some additional Fermi factors \cite{SM}.

Secondly, the divergence is present for more \emph{generic}  quasiparticle dispersions $\epsilon(k)$; the only condition required for a divergence in $\tau^{-1}_k$ is that $\epsilon''(k)\neq 0$. This argument is presented in \cite{SM}; an alternative but less precise argument can be found in the next section. It is moreover present for multi-band models. 

Thirdly, note that Eq.~\eqref{eqn:Stationary phase conditions on shell} is also satisfied when $q,p=0$, but that this does not result in a divergence because the numerator in Eq.~\eqref{eqn:Quasiparticle lifetime lowest order} also vanishes there. This disappearance is due to the Pauli exclusion principle, and is the reason that our divergence has not been observed in continuum spinless Fermi models \cite{bertini_essler}. However, one can evade the Pauli principle by introducing an additional label; in \cite{SM} we confirm that multi-band/spinful systems of Fermions in the continuum exhibit logarithmic divergences at leading order in perturbation theory. Moreover, the Pauli principle is completely defeated if we deal with bosons rather than fermions. The scattering rate for bosons is structurally similar to Eq.~\eqref{eqn:Quasiparticle lifetime lowest order} in the continuum or the lattice, except in the bosonic case the numerator generically does not vanish  at $p,q=0$ so that this point \emph{does} lead to a log-divergence \cite{SM}.

\paragraph{Non-perturbative contributions to Fermi's Golden Rule.---}

We begin by recasting the log-divergence of FGR in real space and time. In this language, the leading order correction to the self-energy is
\begin{equation}\label{eq:self-energy-leading}
    \includegraphics[width=0.8\columnwidth]{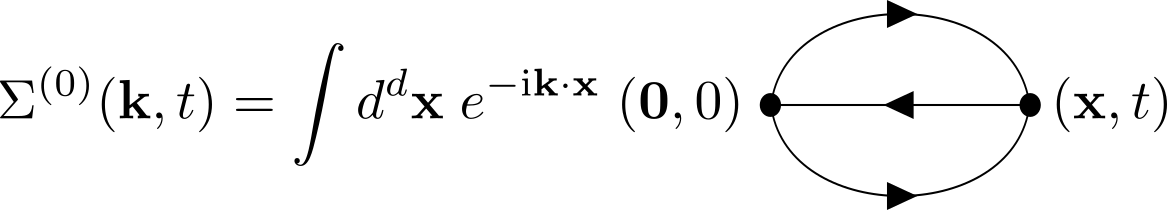}
\end{equation}
The  integrand is a product of three free lattice Feynman propagators, which exhibit Bessel-like behaviour. In $d$-dimensions they scale as  $g(|\boldsymbol{x}|/\sqrt{t}) t^{-d/2}$ within the light-cone, where $g$ is a decaying oscillatory function. Performing the spatial integral with such an ansatz, accounting for both power counting and the rapid fluctuation of the integrand on $\mathcal{O}(\sqrt{t})$ scales, gives ${\Sigma_{k}(t) \sim e^{-\mathrm{i}\epsilon_{k}t}t^{-d}}$. When we Laplace transform this function on-shell in $d=1$ we reproduce the logarithmic divergence Eq.~\eqref{eqn:Quasiparticle lifetime lowest order}. Note that there is no divergence in $d>1$. 

At early times, interactions are unimportant and $\Sigma_{k}(t)$ should decay as $e^{-\mathrm{i}\epsilon_{k}t}/t$ in 1d. Eventually, however,  interactions will dress the propagator lines in Eq.~\eqref{eq:self-energy-leading}, and because $\Sigma_{k}(t)$  is an autocorrelation function \cite{SM} of a non-hydrodynamical operator, we expect it to crossover into a superpolynomial decay  \cite{mcculloch_2025}. The simplest assumption is that the non-interacting approximation breaks down at the characteristic time for quasiparticle scattering $\tau$.  Omitting the explicit momentum dependence here for brevity, we obtain from  Eq.~\eqref{eqn:Quasiparticle decay rate} a self-consistent equation for the lifetime 
\begin{equation}
\label{eqn:Lifetime self consistent equation}
    \frac{1}{\tau} = \Delta^{2}\int^{\tau} \frac{dt}{t} + \Delta^{2}\int_{\tau}^{\infty} dt \; F(t/\tau)
\end{equation}
where the function $F(t)$ decays faster than $1/t$ (likely superpolynomially in practice). Performing the integrals gives $ \tau^{-1} \sim \Delta^{2}\log(\tau)$, which implies that   $ \tau^{-1} \sim \Delta^{2}\log \Delta^{-2}$ in the small $\Delta$ limit, giving our main prediction, Eq.~\eqref{eq:maintauinv}. 
\paragraph{Numerical simulation.---}

\begin{figure*}\label{fig:main}
    \centering
     \includegraphics[width=.99\linewidth]{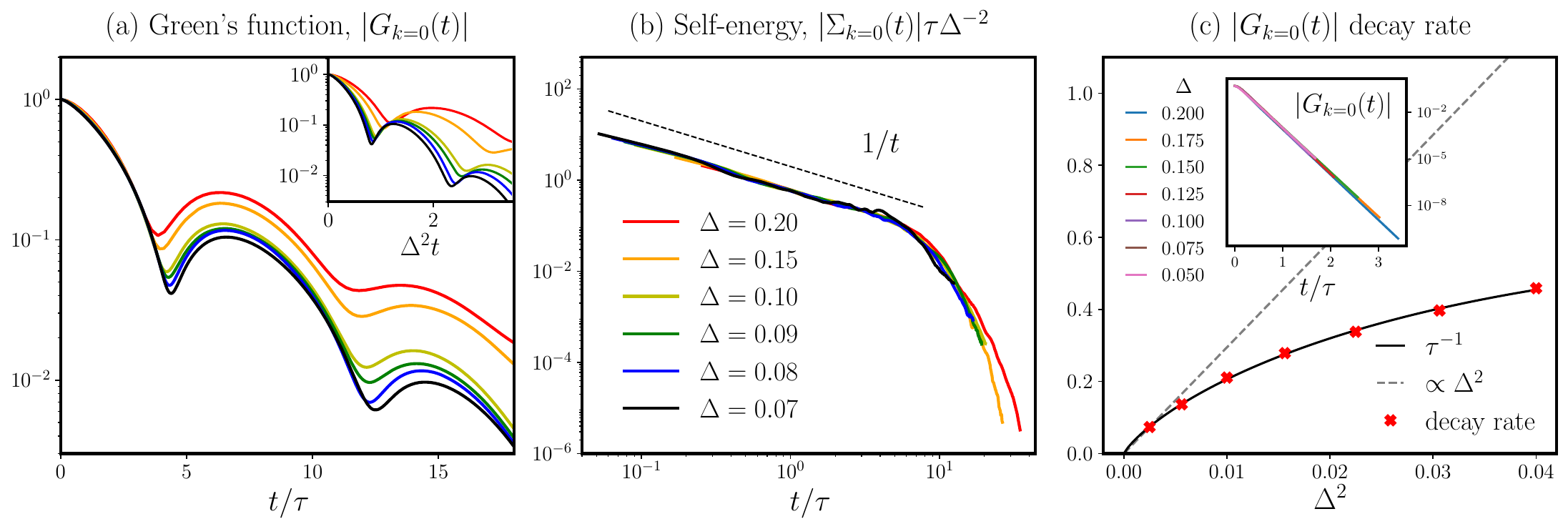}

     \caption{In this figure, $\tau^{-1} \equiv \Delta^{2}\log\Delta^{-2}$ where $\Delta$ is the interaction strength, and the Hamiltonian is Eq.~\eqref{eqn:Hamiltonian}. (a) Decay of the single-particle Green's function with respect to the rescaled time $t/\tau$, for the interaction strengths displayed in (b). We set $k = \pi/L \approx 0$, where $L=600$ is the system size. The Green's function data is extrapolated in $\gamma \to 0$ \cite{SM}. In the inset, we show the same data plotted with respect to the FGR prediction. (b) Single-particle self-energy $\Sigma_k(t)$  vs.~$t/\tau$. The self-energy decays as $|\Sigma_k(t)| \sim 1/t$ up to times $\mathcal{O}(\tau)$, but decays superpolynomially for later times. The scaling collapse is consistent with the logarithmically enhanced decay rate. Here we set $\gamma=0.01$. (c) Data showing the decay rate of $G_{k=0}$ against $\Delta^{2}$ using the melonic approximation (red X-symbols). FGR scaling corresponds to the grey dashed line, while the black solid line is the fit $a\Delta^2\log (b\Delta^{-2})$, with $a\approx 6.7$, $b\approx 0.2$. The data is consistent with the logarithmically corrected prediction, which is further confimed by the scaling collapse of $|G_{k=0}(t)|$  vs.~$t/\tau$ shown in the inset.  \label{fig:GFDAOE}}
\end{figure*}

We perform two numerical checks of the logarithmically-enhanced scattering rate prediction Eq.~\eqref{eq:maintauinv} for the model in Eq.~\eqref{eqn:Hamiltonian}. First, we calculate the single-particle Green's function, $G_k(t)$, which is expected to decay in time with a rate $ \tau^{-1} \sim \Delta^{2}\log\Delta^{-2}$. Second we calculate the real-time self-energy, $\Sigma_k(t)$, which according to Eq.~\eqref{eqn:Lifetime self consistent equation}, should decay as $\sim t^{-1}$ up to a time of order $\tau$, followed by a  superpolynomial decay at later times. 

The simulations are performed in real time, using matrix product state (MPS) methods to evolve operators in the Heisenberg picture. While conventional MPS evolution methods are typically limited to short times (the memory cost scales exponentially with the  total entanglement of the system, which increases linearly under generic unitary time evolution), we overcome this issue using the dissipation-assisted operator evolution algorithm (DAOE)~\cite{DAOE1,jerome_daoe}, which allows us to reach the long timescales needed to verify the scaling correction. We check the numerical convergence in \cite{SM}. 

For the Green's function calculation $G_k(t) = \langle f^{\null}_k(t)  f^\dagger_k(0) \rangle$, we consider the limit $k\to 0$ and infinite temperature. Our results do not change substantially with $k$ except near $k=\pm \pi/2$ (see discussion following Eq.~\ref{eqn:Quasiparticle lifetime ladder series} and \cite{SM}). In Fig.~\ref{fig:GFDAOE}a, we observe a clear decay of the Green's function with a scaling collapse at small $\Delta$ in agreement with the decay rate  $\tau^{-1} = \Delta^{2}\log\Delta^{-2}$. The inset provides a comparison with the scaling collapse if $\tau^{-1} = \Delta^2$, giving a noticeably worse fit. 

The self energy can be written \cite{SM} as 
\begin{equation}\label{eqn:self-energy}
\mathrm{i}\Sigma_k(t) = \lim_{\delta t\to 0}\bra{f^\dagger_k}{\mathcal{U}(\mathcal{Q}\mathcal{U}\mathcal{Q}})^{t/\delta t} \mathcal{U}\ket{f^\dagger_k},
\end{equation}
where $\mathcal{Q}$ projects onto states with at least two fermions, i.e.~$l>1$. Intuitively, Eq.~\eqref{eqn:self-energy} corresponds to the single-particle Greens function projected onto the irreducible diagrams. This form of the self-energy is made amenable to numerical simulation by taking the time-step $\delta t$ small but finite. We then leverage DAOE by making the replacement 
\begin{equation}\label{eq:SEDAOE}
    \mathrm{i}\Sigma_k(t) = \bra{f^\dagger_k}{\mathcal{U}(\mathcal{Q}\mathcal{U}\mathcal{D}\mathcal{Q}})^{t/\delta t} \mathcal{U}\ket{f^\dagger_k}, 
\end{equation}
where now $\mathcal{D}$ is the DAOE dissipation operator. In Fig.~\ref{fig:GFDAOE}b we show the calculated self energies, for different interaction strengths. The data supports our hypothesis for the lifetime in Eq.~(\ref{eqn:Lifetime self consistent equation}), namely that the self-energy decays as $1/t$ until a time of order $\tau$, followed by a superpolynomial decay. The combined results for the self energy and Green's function decay strongly support the suggested correction to Fermi's Golden Rule, Eq.~\eqref{eq:maintauinv}.

\paragraph{Analytical checks.---}
\begin{figure}
    \centering
    \includegraphics[width = 1\linewidth]{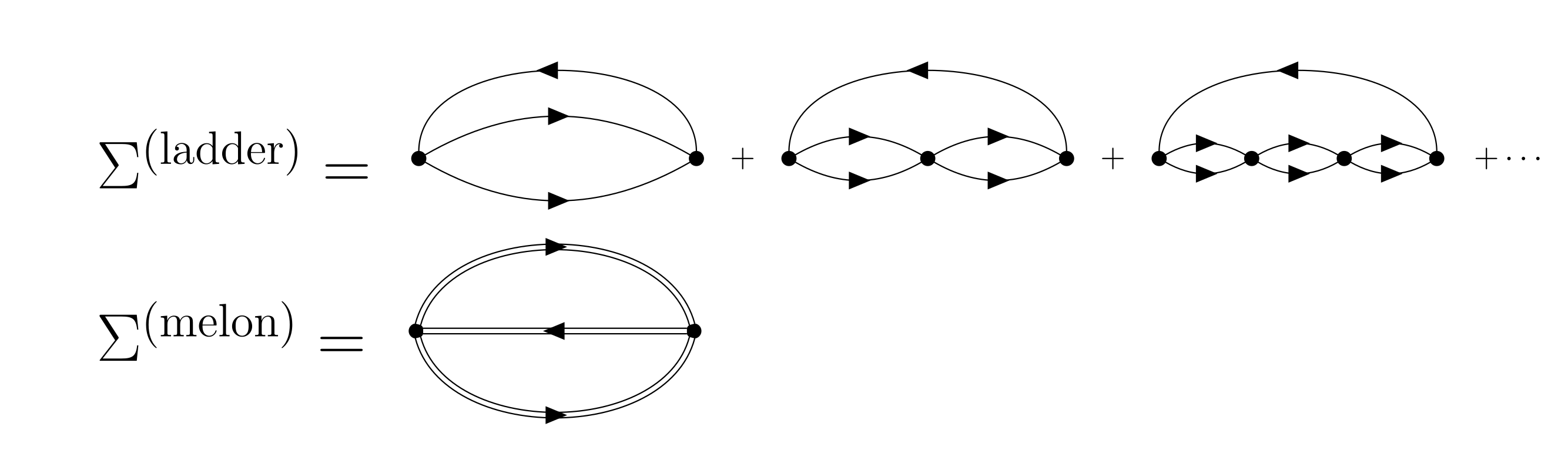}
    \caption{Ladder (above) and melon (below) diagrams that are re-summed to approximate the self-energy. For the melon diagrams the double lines represent interacting propagators $G_{k}$ that are related to the non-interacting propagators $G^{(0)}_{k}$ via $G = G^{(0)}_{k} + G^{(0)}_{k} \Sigma^{(\text{melon})}_{k} G_{k}$.}
    \label{fig:resum}
\end{figure}

We now check our predictions by performing two distinct diagrammatic resummations of the Green's function: the ladder, and the melon sum approximations shown in Fig.~\ref{fig:resum}. 

The former approximates the self-energy with the particle-particle ladder series shown; it can also be done with particle-hole ladders with similar results. 

The ladder re-summation can be evaluated exactly \cite{SM}, leading to expressions for the self-energy and Green's function, and most importantly the quasiparticle lifetime which scales asymptotically as
 \begin{equation}
\label{eqn:Quasiparticle lifetime ladder series}
   \tau^{-1}_{k} \sim |\cos k|^{3}\Delta^{2} \log \Delta^{-2}
   \end{equation}
at small $\Delta$. Note that for $k = \pm \pi/2$ the leading order contribution to the decay rate is $\mathcal{O}(\Delta^{2})$ as mentioned previously.

Our second method for computing the self-energy utilises a melonic resummation implicit in Fig.~\ref{fig:resum}. In practice, we implement it through the memory matrix formalism, which provides an exact equation of motion for the Green's function:
\begin{equation}
\label{eqn:Memory matrix Dyson equation}
    \partial_{t}G_{k} + \mathrm{i}\epsilon_{k}G_{k} + \int_{0}^{t} d\tau\; M_{k}(\tau)G_{k}(t-\tau) = 0,
\end{equation}
where $\epsilon_{k} = -\cos k$ is the single-particle energy, and the memory matrix $M$ is defined in \cite{SM}. The memory matrix $M$ is related to the single-particle self-energy as $M = -\mathrm{i}\Sigma$. In real time, recall that the self-energy can be thought of as an auto-correlation of three fermionic operators. In the melonic approximation, we approximate this autocorrelator with a product of three factors of the full propagator $G_k(t)$. 

We find a numerical solution to Eq.~\ref{eqn:Memory matrix Dyson equation} and extract the quasiparticle decay rate from the long time exponential decay of $G_{k}$. The scaling of this rate with the interaction strength $\Delta$ is far more consistent with our predicted logarithmically enhanced decay rate Eq.~\eqref{eq:maintauinv} as opposed to the naive FGR prediction of $\Delta^{2}$, as is show in Fig.~\ref{fig:GFDAOE}c. 

The ladder and melon approximations are very different in character. The ladder resummation results in a self-energy that decays as a power law ($t^{-2}$) decay, which is substantially slower than the superpolynomial decay generally expected. The melon approximation gives a clear exponential temporal decay of the self-energy. Nevertheless, both analytical methods agree with the prediction $ \tau^{-1} \sim \Delta^{2}\log\Delta^{-2}$ for almost all values of $k$.

Note that at exceptional values of  $k = \pm \pi/2$, the logarithmic enhancement in Eq.~\eqref{eqn:Quasiparticle lifetime ladder series} disappears. This corresponds to the fact that the FGR result does \emph{not} diverge at these wavevectors because the two solutions to Eq.~\eqref{eqn:Stationary phase conditions on shell} merge. This result is consistent with our numerics for the DAOE which also show the absence of logarithmic enhancement at these wavevectors \cite{SM}.

\paragraph{Other models. ---}
Thus far we have studied the quasiparticle lifetime in a model that happens to be integrable.
For completeness we also perform DAOE numerics for the spinless fermion model Eq.~\eqref{eqn:Hamiltonian} with a staggered chemical potential, $H \to H + h\sum_{j}(-1)^{j}f_{j}^{\dagger}f_j^{\null}$  \cite{SM}. For this non-integrable model \cite{staggeredxxznotintegrable} the characteristic decay timescale for the single-particle Green's function/self-energy is again consistent with our non-perturbative prediction. Furthermore, the melonic approximation to the staggered model gives the same final scaling result for the decay rate \cite{SM}.

The next example demonstrates that the logarithmic enhancement can also appear in classical systems. Consider a classical configuration space with fields $\psi_{x}$ on each site obeying the classical Poisson bracket relation $\left\{\psi_{x}^{*},\psi_{y}\right\} = \mathrm{i}\delta_{xy}$; this symplectic space is the classical limit of the bosonic Fock space on the lattice. Our evolution is Floquet, consisting of alternating free/interacting steps
\begin{equation}
\label{eq:classical}
\begin{split}
    \psi_{x}(t+1) &= e^{\mathrm{i}\Delta\left\{\sum_{r}|\psi_{r}|^{4},\bullet\right\}}e^{\mathrm{i}\left\{H_{0},\bullet\right\}}\psi_{x}(t),
\end{split}
\end{equation}
where $H_{0} = \sum_{rr'}(h_{0})_{rr'} \psi_{r}^{*}\psi^{\null}_{r'}$ is a quadratic hermitian Hamiltonian and the interactions are of the on-site density-density type. The correlation function $C_{k}(t) \propto \sum_{x}e^{-\mathrm{i}kx}\langle \psi^{\null}_{x}(t)\psi_{0}^{*}(0)\rangle$ is the classical analogue of the single-particle Green's function Eq.~\eqref{eqn:single-particle Green's function}. We find \cite{SM} that $C_{k}$ exhibits long-time exponential decay with a rate that is again much better fit by $\Delta^{2} \log \Delta^{-2}$ than by $\Delta^{2}$. We expect the logarithmic enhancement to be generic in similar classical models: the key condition is that in the absence of interactions/non-linearities the model has normal-modes with a sufficiently generic `Schr\"odinger' dispersion ${\partial_t =- \mathrm{i} \epsilon_k}$.

Given that the log-enhancement appears in a classical model, is it truly a quantum effect? Our answer is a qualified yes. In the context of systems of quantum particles, the log-enhancement requires that free particles are subject to the quantum coherent effects that give rise to the Schr\"odinger propagator. The effect should disappear in a classical/noisy environment, e.g., in the presence of extrinsic dephasing. 

\paragraph{Discussion.---}
We predict that the quasiparticle decay rate for a wide class of weakly-interacting models is logarithmically-enhanced in $d=1$, i.e., $\tau^{-1} \sim \Delta^{2} \log \Delta^{-2}$. We confirmed this scaling with DAOE numerical simulations and two different (ladder, and melonic) semi-analytic resummations.

There are numerous future directions. It is a priority to confirm our scaling prediction analytically, perhaps in an integrable model, e.g., XXZ at small $\Delta$, or Lieb-Liniger in the weakly-interacting limit. While there are many exact results in the field, there are none at present for the single-particle spectral function, mainly because it does not fit into the generalized hydrodynamics (GHD) framework.  All of the models studied here have either continuous or lattice translation symmetry; it is worth determining how disorder affects the log-enhancement.

It is important to clarify whether the logarithmic enhancement appears in bulk transport coefficients. Naive scaling arguments suggest that the charge diffusion constant is proportional to the scattering time, $D = v^{2}\tau $, so that the diffusion constant is logarithmically suppressed  $D\sim v^2/( \Delta^2\log\Delta^{-2})$ compared to the FGR prediction.  Previous numerical results \cite{jerome_daoe} give an anomalously small prefactor when fitting the diffusion constant with an FGR scaling. The naive prediction of a logarithmically suppressed diffusion constant would go some way to account for this result. However the formula $D=v^2 \tau$ is too naive in certain cases \cite{mahan2013many}, because it does not account for whether the dominant quasiparticle scattering processes relax the current effectively. For the model in Eq.~\eqref{eqn:Hamiltonian} (or rather its non-integrable version, which is expected to show late-time diffusive behaviour), the divergent scattering process does not change the group velocities, hence it is unclear whether the log-enhancement is visible in the transport coefficient. Further numerical and analytical studies are needed. 

Lastly, we turn to the question of experimental realisations. The single-particle spectral function, our central object of study, can be probed with ARPES in the solid-state \cite{sobota2020electronic}. While many recent experiments in 1d have focused on probing the low-temperature Luttinger liquid regime \cite{bouchoule2025glance, blumenstein2011atomically, ohtsubo2015surface}, our results should apply in many of the same systems at higher temperatures, assuming weak interactions and moreover a weak coupling to auxiliary degrees of freedom (e.g., phonons). Analogous `spectroscopic' schemes exist in cold-atomic systems \cite{Jin_spectral_function,Carusotto_spectral_function, brown2020angle}; said systems are particularly promising, allowing one to engineer the clean and closed 1d systems for which we predict the logarithmic enhancement. Experiments realising cold atomic fermions with tunable spin and interaction strength have been realised \cite{pagano2014one}, which would be a natural setting to test our prediction for multi-band fermion models. Finally, measuring the charge diffusion constant in 1d non-integrable but tunably weakly interacting system of bosons or fermions \cite{tunable-two-species-fermi-hubbard,Wienand_2024} would clarify the link between quasiparticle lifetime and transport coefficients.

\paragraph{Acknowledgements.---} We thank F. Essler, L. Glazman, E. McCulloch, D. Huse and C. Berthod for useful conversations.  T.Y.  is supported by EPSRC studentship. C.K. was supported by a UKRI FLF through MR/T040947/2 and MR/Z000297/1.
\bibliography{main}

\pagebreak

\widetext

\newpage

\makeatletter
\begin{center}
\textbf{\large Supplemental Materials -- Breakdown of Fermi's golden rule in 1d systems at non-zero temperature}
\end{center}

\author{Thomas Young} 
  \affiliation{Department of Physics, King’s College London, Strand WC2R 2LS, UK}

\author{Jerome Lloyd}
\affiliation{Department of Theoretical Physics, University of Geneva, Geneva, Switzerland}

\author{Curt von Keyserlingk} 
 \affiliation{Department of Physics, King’s College London, Strand WC2R 2LS, UK}

\date{\today}
\maketitle

\section{Arguing that the divergence is generic}\label{app:generic}
In this section we show that the logarithmic divergence is generic in translation invariant bosonic and fermionic 1d systems, both on the lattice and in the continuum. There are a few caveats which we spell out below.

\subsection{Fermions on the lattice}
In this section we argue that the logarithmic divergence in FGR occurs for generic translation invariant models of 1d fermions on the lattice. We show that it persists for finite nonzero temperatures, and also for more generic dispersions. 
\subsubsection{Finite temperature}
We compute the FGR  quasiparticle lifetime at finite temperature $\beta^{-1}$ using the imaginary time formalism, and later analytically continuing to real-time/frequency. For a single-band model, the leading-order (FGR) contribution to the self-energy takes the form
\begin{equation}
\Sigma_{k}(\tau) \approx 2\int \frac{dqdp}{(2\pi)^{2}} \;v(q,p)^{2} G_{\text{p}}^{(0)}(q+k,\tau)G_{\text{p}}^{(0)}(p+k,\tau)G_{\text{h}}^{(0)}(q+p+k,\tau),
\end{equation}
where $G_{\text{p,h}}^{(0)}(k,\tau)$ are imaginary-time non-interacting particle/hole Green's functions, and the vertex factor $v(q,p)$ encodes the form of the interactions. For a single-band model $v(q,p)$  vanishes for $q=p$ because of the Pauli-exclusion principle; it is equal to $ \Delta(\cos q - \cos p)$ for the model Eq.~(2) in the main text. The self-energy can be written in terms of its Fourier components which are non-zero for fermionic Matsubara freqencies $\omega_{n} = \frac{(2n+1)\pi}{\beta}$
\begin{equation}
    \Sigma_{k}(i\omega_{n}) = -2\int \frac{dqdp}{(2\pi)^{2}} v(q,p)^{2}\; \frac{n_{k+q+p}(n_{k+q}+n_{k+p}-1) - n_{k+q}n_{k+p}}{i\omega_{n} - \epsilon_{k+q}-\epsilon_{k+p}+\epsilon_{k+q+p}},
\end{equation}
where $n_{k}$ are Fermi-functions. Performing the analytic continuation $i\omega_{n} \to \omega + i0^{+}$ and taking the on-shell limit $\omega \to \epsilon_{k}$ yields the following FGR expression for the quasiparticle lifetime
\begin{equation}\label{eq:FGRfermionsfiniteT}
    \frac{1}{\tau_{k}} = 2\;\text{Im} \int \frac{dqdp}{(2\pi)^{2}} v(q,p)^{2}\; \frac{n_{k+q+p}(n_{k+q}+n_{k+p}-1) - n_{k+q}n_{k+p}}{\epsilon_{k} - \epsilon_{k+q}-\epsilon_{k+p}+\epsilon_{k+q+p}+i0^{+}}.
\end{equation}
As discussed near Eq.~\eqref{eqn:Quasiparticle lifetime lowest order}, there are logarithmic divergences associated with special points in $(q,p)$ space; for the cosine dispersion these points are at $(q_{*},p_{*}) = (0,\pi-2k),(\pi-2k,0)$. The logarithmic divergences are  accompanied by a temperature dependent pre-factor $\propto n_{k+p_{*}}(1-n_{k+p_{*}})$, which is non-zero at $T>0$. Therefore the logarithmic divergence persists for all non-zero temperatures. 

If we consider the zero-temperature ($\beta \to \infty$) limit we notice that the prefactor $n_{k+p_{*}}(1-n_{k+p_{*}})$ vanishes for all $k$ except when $k+p_{*}$ is on the Fermi-surface ($\pm k_{F})$. For the weakly interacting Eq.~(2) model ($\epsilon_{k} = -\cos(k), v(q,p) = \Delta (\cos q - \cos p)$) the log-divergence in the FGR formula for $\tau_{k}^{-1}$ survives only for $k = \pi -k_{F}$ which is consistent with the explicit calculation at $T=0$ \cite{Pereira_2009}, however at zero temperature this divergence is in fact nullified by Luttinger liquid physics resulting in an infinite lifetime for these particlular quasiparticles.

\subsubsection{General dispersion}
The log divergence occurs for generic dispersions. Consider the expression appearing in the denominator of the FGR integral Eq.~\eqref{eq:FGRfermionsfiniteT}
\begin{equation}
\phi_{k}(q,p)\equiv\epsilon_{k}+\epsilon_{q+p+k}-\epsilon_{k+q}-\epsilon_{k+p}.
\end{equation}
The log divergence requires that there exist $(q_{*},p_{*})$ such that
\begin{align}
\phi_{k}(q_{*},p_{*}) & =0\label{eq:on-shell}\\
\nabla  \phi_{k} \mid_{q=q_{*},p=p_{*}} & =0\label{eq:grp-velocity}.
\end{align} 
This ensures that the denominator in the above integral vanishes sufficiently quickly at the point $(q_{*},p_{*})$ to give a log-divergence. The first condition is just the requirement that the process is on-shell. The second condition can be written explicitly as

\begin{align*}
 & \partial_{q}\epsilon_{q+p+k}-\partial_{q}\epsilon_{k+q}=0\\
 & \partial_{p}\epsilon_{q+p+k}-\partial_{p}\epsilon_{k+p}=0.
\end{align*}
Together these imply that log divergences are associated with processes where the outgoing particle and particle-hole pairs move together with the same group velocity.

For the standard nearest neighbour hopping dispersion, there are two log divergences that contribute equally by symmetry: $(q_{*},p_{*})=(0,\pi-2k)$ and $(q_{*},p_{*}) =(\pi-2k,0)$ respectively. But the log-divergences seem to occur for even more general dispersions. For example, if we consider the line $p=0$, then we are automatically on-shell $\phi=0$. The first group velocity condition is automatically satisfied, and we only need to satisfy the second, which amounts to finding a $q$ such that 

\begin{equation}
\label{eq:Velocity matching single band}
\epsilon'(q+k)-\epsilon'(k)=0
\end{equation}
Except at exceptional $k$ points, this tends to have least two solutions for general smooth dispersions: $q=0$, and $q=K\in(0,2\pi)$. The existence of a second solution follows from the periodicity of $\epsilon'$ and the mild requirement that $\epsilon''(k)\neq0$.

The $q=p=0$ point is not interesting for fermions on the lattice; the log divergence is mollified by the vertex $v$ factor in Eq.~\eqref{eq:FGRfermionsfiniteT} which vanishes at $q=p$. The second solution $q_*=K,p_*=0$ (and likewise $q_*=0,p_*=K$) \emph{does} generically contribute a log divergence. The existence of this second solution is tied to the periodicity of $\epsilon(k)$ which in turn is tied to the fact we are on the lattice and not in the continuum.

\subsubsection{Multi-band/spinful Models}For multi-band models there are multiple diagrams at second order in $\Delta$ contributing to the FGR formula for the quasiparticle decay rate. It is important to note that each such contribution is non-negative, hence there are no cancellations between diagrams.  

We consider a general weakly interacting model with quartic U(1) conserving interactions. Denote the single particle dispersions for particles in band $b$ with momenta $k$ with $\epsilon_{b}(k)$. Charge conservation implies that a particle scatters into a particle-particle-hole triplet. A general on-shell self-energy diagram at second order in $\Delta$, for a particle of momenta $k$ in band $b$, will have the following expression in the denominator of the integrand (see Eq.~\eqref{eq:FGRfermionsfiniteT}) 
\begin{equation}
    \phi_{k}(q,p) = \epsilon_{b}(k) + \epsilon_{b_{1}}(k+q+p) - \epsilon_{b_{2}}(k+q) - \epsilon_{b_{3}}(k+p).
\end{equation}
In order for this diagram to have a log-divergent contribution to the FGR decay rate we require 
\begin{equation}
\label{eq:Log divergence condition}
    \phi_{k} = \partial_{q}\phi_{k} = \partial_{p}\phi_{k} = 0
\end{equation}
at some point $(q,p) = (q_{*},p_{*})$. In general the different band dispersions $\epsilon_{b}(k)$ are not related to one another. There is the potential for a divergence if we examine the case $b_{2} = b,b_{1} = b_{3}=b'$
\begin{equation}
    \phi_{k}(q,p) = \big(\epsilon_{b}(k) - \epsilon_{b}(k+q)\big) + \big(\epsilon_{b'}(k+q+p) - \epsilon_{b'}(k+p)\big)
\end{equation}
from which we see that the on-shell condition, as well as one of the group velocity conditions in Eq. ~\eqref{eq:grp-velocity} are satisfied for $q_{*} = 0$. The last remaining condition is
\begin{equation}
\label{eq:Velocity matching multi band}
    \epsilon'_{b'}(k+p_{*}) = \epsilon'_{b}(k).
\end{equation}
We have previously argued that for $b=b'$ there are at least two solutions to this equation (provided $\epsilon''_{b}(k)\neq 0$) for which all but the trivial $p_{*} = 0$ solution will contribute a log-divergence (since the $p_{*} = 0$ divergence is nullified by Pauli exclusion). For $b' \neq b$ are only guaranteed a solution for all $k,b$ if the group velocities in each band have the same upper/lower bounds, i.e., $\text{max}|\epsilon'_{b}(k)| = c$ for each band $b$, where $c$ is some constant. In summary, multiband fermion models on the lattice generically have the logarithmic divergence for each flavor of fermion. They can also have further logarithmic divergences coming from interflavor interactions, but in this case the existence of the logarithmic divergence will depend on the details of the band and the incoming wavevector $k$.

We remark that the two-band model studied in \cite{Coulomb_drag_2019} (Fermi-Hubbard model) permits only scattering between the two bands (i.e., the model has only an $n_\uparrow n_\downarrow$ interaction). But as the two bands have identical dispersions $\epsilon_{\uparrow}(k) = \epsilon_{\downarrow}(k)$, their group velocities have the same upper/lower bounds, and hence the FGR decay rate is log-divergent. 

\subsection{Bosons on the lattice}
In this section we argue that the logarithmic divergence for FGR also arises for 1d Bosons. It is in some sense more prevalent since Bosons do not obey the Pauli-exclusion principle, which can sometimes nullify the divergence. At finite temperature the self-energy can be expressed in terms of its Fourier components, which are non-zero for bosonic Matsubara frequencies $\omega_{m} = \frac{2m}{\beta}\pi$ 
\begin{equation}
    \Sigma_{k}(i\omega_{m}) = 2\int \frac{dqdp}{(2\pi)^{2}}\; v(q,p)^{2} \frac{n_{q+p+k}(1+n_{q+k}+n_{p+k}) - n_{q+k}n_{p+k}}{i\omega_{m}+\epsilon_{q+p+k}-\epsilon_{q+k}-\epsilon_{p+k}},
\end{equation}
where $n_{k}$ are Bose occupation functions 
\begin{equation}
    n_{k} = \frac{1}{e^{\beta(\epsilon_{k}-\mu)}-1}.
\end{equation}
We consider a sufficiently negative chemical potential $\mu$ such that the Bose function is finite for all momenta $k$. Performing the analytic continuation $i\omega_{m} \to \omega + i0^{+}$ and taking the on-shell limit $\omega \to \epsilon_{k}$ yields the following FGR expression for the quasiparticle lifetime
\begin{equation}
\label{eq:FiniteTbosonsFGR}
    \frac{1}{\tau_{k}} = -2\;\text{Im} \int \frac{dqdp}{(2\pi)^{2}}\; v(q,p)^{2} \frac{n_{q+p+k}(1+n_{q+k}+n_{p+k}) - n_{q+k}n_{p+k}}{\epsilon_{k}+\epsilon_{q+p+k}-\epsilon_{q+k}-\epsilon_{p+k} + i0^{+}},
\end{equation}
for which there are  the same divergences as for the finite temperature fermion models, accompanied this time by a factor $n_{k+p_{*}}(1+n_{k+p_{*}})$. Additionally, the logarithmic divergence at $q=p=0$ is no longer nullified by Pauli exclusion. Therefore there is an additional logarithmic divergence with a prefactor $n_{k}(1+n_{k})$. 
\subsection{Continuum models}
The log-divergence persists for bosons in the continuum. This is apparent from Eq.~\eqref{eq:FiniteTbosonsFGR}, which turns into an infinite integral with a generally non-periodic/quadratic dispersion. However, neither of these modifications changes the logarithmic nature of the singularity at $p,q=0$, provided $T>0$. 

For Fermions in the continuum, the divergence persists for generic \emph{multi-}band models. We consider $T<\infty$ only to ensure UV convergence of the momentum integrals. The FGR scattering rate is the same as Eq.~\eqref{eq:FGRfermionsfiniteT}, except once again the momentum integrals are over $\mathbb{R}^2$ and the dispersions are generically unbounded with wave-vector.

For a single band fermion model, there is no logarithmic divergence. That is because the only potential divergent point $p=q=0$ is nullified by the disappearance of $v(q,p)$ at that point \cite{bertini_essler}. Note the other potential divergent points e.g., $p=\pi-2 k,q=0$, do not appear in the continuum as they require a periodic dispersion.  

Multi-band fermion models, on the other hand, do generically possess a log divergence in the continuum. Recall that such a divergence requires satisfying the on-shell and two velocity matching conditions Eq.~\eqref{eq:Log divergence condition}. Two of these conditions are satisfied by setting $q=0$. The remaining condition,  Eq.~\eqref{eq:Velocity matching multi band}, can be satisfied for any $k$ by choosing $b \neq b'$ and an appropriate $p=p_*$; the existence of such a $p_{*}$ is guaranteed by the mild condition that the range of the derivatives $\epsilon'_b(k)$  is $(-\infty,\infty)$. In particular, the 2-band continuum Fermi-Hubbard model should have a logarithmic divergence.

\section{Memory matrix formalism}
This appendix introduces the memory matrix formalism for a model of
spinless Fermions in a staggered field. This provides the framework
for our non-perturbative melonic approximation
to the Green's function. We work at infinite temperature and half-filling,
which is reflected in our choice of operator inner product $\langle O|O'\rangle=\mathrm{tr}(O^{\dagger}O')/\mathrm{tr}(I)$. We start by defining the single particle Green's function 
\begin{equation}
    G_{AB}(t) \equiv \langle f_{A}^{\dagger}(t)|f_{B}^{\dagger}\rangle = \langle f_{A}^{\dagger}| e^{-i\mathcal{L}t}|f_{B}^{\dagger}\rangle,
\end{equation}
where $A,B$ may be real or Fourier-space indices and $\mathcal{L} = \left[H,\cdot\right]$. We define the fast/slow projectors $P,Q$ 
\begin{equation}
    P = \sum_{A} \frac{|f_{A}^{\dagger}\rangle \langle f_{A}^{\dagger}|}{\langle f_{A}^{\dagger}|f_{A}^{\dagger}\rangle}
\end{equation}
and $Q = I - P$, i.e., $Q$ projects out components containing more than one fermion operator, where we assume that $\langle f_A|f_B\rangle \propto \delta_{AB}$. Taking the Laplace transform of the single particle Green's function 
\begin{equation}
    G_{AB}(z) \equiv \int_{0}^{\infty}dt\; e^{izt} G_{AB}(t) = i \langle f_{A}^{\dagger}| \left(z - \mathcal{L}\right)^{-1}|f_{B}^{\dagger}\rangle.
\end{equation}
In the following we will make use of the operator identity 
\begin{equation}
 (X+Y)^{-1} = X^{-1} - X^{-1}Y(X+Y)^{-1}
\end{equation}
firstly to re-write the Green's function using the fast/slow projectors 
\begin{equation}
    G_{AB}(z) = \frac{i}{z}\langle f_{A}^{\dagger}|f_{B}^{\dagger}\rangle + i\sum_{C} \frac{1}{\langle f_{C}^{\dagger}|f_{C}^{\dagger}\rangle} \langle f_{A}^{\dagger}|(z- \mathcal{L}Q)^{-1}\mathcal{L}|f_{C}^{\dagger}\rangle \langle f_{C}^{\dagger}|(z-\mathcal{L})^{-1}|f_{B}^{\dagger}\rangle 
\end{equation}
in doing so we have used the fact that $(z-\mathcal{L}Q)^{-1}|f_{A}^{\dagger}\rangle = \frac{1}{z}|f_{A}^{\dagger}\rangle$.  Defining $\chi_{AB} \equiv \langle f_{A}^{\dagger}|f_{B}^{\dagger}\rangle = \frac{1}{2}\delta_{AB}$ and performing a further decomposition of $(z-\mathcal{L}Q)^{-1}$ yields 
\begin{align*}
    G_{AB}(z) &= \frac{i}{z}\chi_{AB} + \frac{1}{z} \sum_{C,D} \langle f_{A}^{\dagger}|\left( 1 + \mathcal{L}Q(z - \mathcal{L}Q)^{-1}\right)\mathcal{L}|f_{C}^{\dagger}\rangle \chi^{-1}_{CD} G_{DB}(z) \\
    &= \frac{i}{z}\chi_{AB} + \frac{1}{z} \sum_{C,D} \langle f_{A}^{\dagger}|\mathcal{L}|f_{C}^{\dagger}\rangle \chi^{-1}_{CD} G_{DB}(z) + \frac{1}{z} \sum_{C,D}\langle f_{A}^{\dagger} |\mathcal{L}Q (z - Q\mathcal{L}Q)^{-1}Q\mathcal{L} | f_{C}^{\dagger}\rangle \chi_{CD}^{-1}G_{DB}(z).
\end{align*}
We define the `Memory matrix' $M(z)$, and the matrix $\epsilon$ via 
\begin{equation}
\begin{split}
    (M\chi)_{AB}(z) &= i \langle f_{A}^{\dagger}|\mathcal{L}Q(z-Q\mathcal{L}Q)^{-1}Q\mathcal{L}|f_{B}^{\dagger} \rangle \\
    (\epsilon \chi)_{AB} &= \langle f_{A}^{\dagger}|\mathcal{L}|f_{B}^{\dagger}\rangle 
\end{split}
\label{eqn:Memory matrix laplace transform}
\end{equation}
resulting in the more familiar matrix equation for the Green's function 
\begin{equation}
    \mathbf{G}(z) = i\left(z\mathbf{I} - \boldsymbol{\epsilon} + i\mathbf{M}(z)\right)^{-1} \boldsymbol{\chi}.
\label{eqn:Green's function laplace tranasform}
\end{equation}
At this point we compare our expression for the single particle Green's function to those more commonly found in the literature and notice the following
\begin{itemize}
    \item The matrix $\boldsymbol{\epsilon}$ encodes the single particle energy but additionally includes the leading order diagram in perturbation theory (Hartree term). 
    \item At least at infinite temperature, the memory matrix  can be identified with the standard retarded self-energy $\mathbf{M} = \mathrm{i} \boldsymbol{\Sigma}^{(A)}$ \cite{Kamenev_2011}.
\end{itemize}
From (\ref{eqn:Memory matrix laplace transform}) we infer that the real-time self-energy takes the form 
\begin{equation}\label{eq:self-energy-mem}
    (M\chi)_{AB}(t) = \langle f_{A}^{\dagger}|\mathcal{L}Qe^{-iQ\mathcal{L}Qt}Q\mathcal{L}|f_{B}^{\dagger}\rangle.
\end{equation}
Finally, we remark the Dyson equation (\ref{eqn:Green's function laplace tranasform}) is precisely the Laplace transform of 
\begin{equation}
    \partial_{t} \mathbf{G}(t) + i\boldsymbol{\epsilon}\, \mathbf{G}(t) + \int_{0}^{t} dt' \; \mathbf{M}(t')\mathbf{G}(t-t') = 0
\label{eqn:Dyson equation real time}
\end{equation}
using $\boldsymbol{\chi} = \mathbf{G}(t=0)$.

\subsection{Memory matrix calculation for the staggered XXZ model}
Throughout the main text we work with the staggered field XXZ model (in both its integrable and non-integrable limits), with Hamiltonian 
\begin{equation}
    H = -\sum_{r} \left(\frac{1}{2}(f_{r+1}^{\dagger}f_{r} + \text{h.c}) + h(-1)^{r}(1 - 2f_{r}^{\dagger}f_{r})\right) + \Delta \sum_{r}(1-2f_{r}^{\dagger}f_{r})(1 - 2f_{r+1}^{\dagger}f_{r+1})
\label{eqn:Staggered field Hamiltonian}
\end{equation}
that reduces to the (integrable) XXZ model for $h=0$. We work in
periodic boundary conditions with $L=2N$ sites. Let $x$ label unit
cells, and $\eta=\pm$ label the first (even) and second (odd) site
in the unit cell so that $f_{x,\eta}=f_{2x+(1-\eta)/2}$. Furthermore
let $f_{k,\eta}=N^{-1/2}\sum_{x}e^{-\mathrm{i}kx}f_{x,\eta}$ so that

\begin{align*}
H & =\sum_{k}\left[\begin{array}{cc}
f_{k,+}^{\dagger} & f_{k,-}^{\dagger}\end{array}\right]\underbrace{\left[\begin{array}{cc}
2h & -(1/2)(1+e^{\mathrm{-i}k})\\
-(1/2)(1+e^{\mathrm{+i}k}) & -2h
\end{array}\right]}_{\boldsymbol{\epsilon}_{k}}\left[\begin{array}{c}
f_{k,+}\\
f_{k,-}
\end{array}\right]\\
 & +\Delta\sum_{x}(1-2f_{x,+}^{\dagger}f_{x,+})(1-2f_{x,-}^{\dagger}f_{x,-})+\Delta\sum_{x}(1-2f_{x,-}^{\dagger}f_{x,-})(1-2f_{x+1,+}^{\dagger}f_{x+1,+}).
\end{align*}
We will study the decay of
the single particle propagators

\begin{align*}
G_{k,\eta\eta'}(t) & \equiv\langle f_{k,\eta}^{\dagger}|e^{-\mathrm{i}\mathcal{L}t}|f_{k,\eta'}^{\dagger}\rangle\\
G_{k,\eta\eta'}^{(-)}(t) & \equiv\langle f_{k,\eta}|e^{-\mathrm{i}\mathcal{L}t}|f_{k,\eta'}\rangle\\
 & =[G_{k,\eta\eta'}(t)]^{*}
\end{align*}
where $\mathcal{L}=[H,\cdot]$. Note the last equation does not hold
at more general temperatures, where instead we find $G_{k,\eta\eta'}^{(-)}(t)=[G_{k,\eta\eta'}(t-\mathrm{i}\beta)]^{*}$.
In the absence of interactions ($\Delta=0$) we denote the propagators
by $G_{k,\eta\eta'}^{(0)}$ and it is straightforward to show that
the non-interacting propagator is

\begin{align*}
G_{k,\eta\eta'}^{(0)}(t) & =\frac{1}{2}(e^{-i\boldsymbol{\epsilon}_{k}t})_{\eta,\eta'}.
\end{align*}
Explicitly

\begin{align*}
G_{k,\eta\eta'}^{(0)}(t) & =\frac{1}{2}\left(\begin{array}{cc}
\cos\left(\omega_{k}t\right)-\frac{2ih\sin\left(\omega_{k}t\right)}{\omega_{k}} & \frac{i\left(1+e^{-ik}\right)\sin\left(\omega_{k}t\right)}{2\omega_{k}}\\
\frac{i\left(1+e^{ik}\right)\sin\left(\omega_{k}t\right)}{2\omega_{k}} & \cos\left(\omega_{k}t\right)+\frac{2ih\sin\left(\omega_{k}t\right)}{\omega_{k}}
\end{array}\right)_{\eta\eta'},
\end{align*}
where $\omega_{k}=\sqrt{\frac{1+\cos k}{2}+4h^{2}}$. 

We recall the definition of the Memory matrix, which in real space takes the form 
\begin{equation} 
    (M\chi)_{rr'}(t) = \langle f_{r}^{\dagger}| \mathcal{L}Qe^{-iQ\mathcal{L}Qt} Q\mathcal{L}|f_{r'}^{\dagger}\rangle.
\end{equation}
Performing the non-interacting (FGR) type approximation for $M$ amounts to replacing the time evolution with the non-interacting one 
\begin{equation}
    (M\chi)_{rr'}^{(0)}(t) = \langle f_{r}^{\dagger}| \mathcal{L}Qe^{-iQ\mathcal{L}_{0}Qt} Q\mathcal{L}|f_{r'}^{\dagger}\rangle
\end{equation}
where $\mathcal{L}_{0} = \left[H_{0},\cdot\right]$ with $H_{0}$ being the Hamiltonian (\ref{eqn:Staggered field Hamiltonian}) in the limit $\Delta= 0$. Noting that $\left[ \sum_{r'}n_{r'}n_{r'+1},f_{r}^{\dagger}\right] = (n_{r+1} + n_{r-1})f_{r}^{\dagger}$ where $n_{r} = f_{r}^{\dagger}f_{r}$, we obtain the following 
\begin{align*}
    (M\chi)_{rr'}^{(0)} &= 16\Delta^{2}\sum_{\delta,\delta' = \pm1} \langle n_{r+\delta}f_{r}^{\dagger}| e^{-iQ\mathcal{L}_{0}Qt}|n_{r'+\delta'}f_{r'}^{\dagger}\rangle \\
    &= 16\Delta^{2} \sum_{\delta,\delta'} \left[G_{r+\delta,r'+\delta'}^{(0)}(t)\right]^{*}\left(G_{r+\delta,r'+\delta'}^{(0}(t)G_{r,r'}^{(0)}(t) - G_{r+\delta,r'}^{(0)}(t)G_{r,r'+\delta'}^{(0)}(t)\right)
\end{align*}
where in the last line we have used Wick's theorem. A straightforward but tedious calculation yields the Memory matrix in momentum space 
\begin{align}
(M\chi)_{k,\eta,\eta'}^{(0)} & =\frac{64\Delta^{2}}{N^{2}}\sum_{p_{1}p_{2}}\cos^{2}(p_{2}/2)e^{ip_{2}(a'-a)}G_{k+p_{1},-\eta,-\eta'}^{(0)}\left[G_{k+p_{1}+p_{2},-\eta,-\eta'}^{(0)}\right]^{*}G_{k+p_{2},\eta,\eta'}^{(0)}\nonumber \\
 & -\frac{16\Delta^{2}}{N^{2}}\sum_{p_{1}p_{2}}(1+e^{ip_{2}})(1+e^{-ip_{1}})\times e^{i(-p_{2})(a)}e^{i(p_{1})a'}[G_{k+p_{1},-\eta,\eta'}^{(0)}][G_{k+p_{1}+p_{2},-\eta,-\eta'}^{(0)}]^{*}[G_{k+p_{2},\eta,-\eta'}^{(0)}]\label{eq:M-non-interacting}
\end{align}
where $a = (1-\eta)/2$ and similarly for $a'$. Note the in the $h=0$ (integrable) limit, this expression reduces to 
\begin{equation}
    M_{k}^{(0)}(t) = \frac{8\Delta^{2}}{L^{2}} \sum_{p,q} (\cos q - \cos p)^{2}e^{i(\epsilon_{k+q+p} - \epsilon_{k+q} - \epsilon_{k+p})t}
\end{equation}
which is consistent with the FGR approximation for the quasiparticle lifetime that appears in the main text. 
\subsection{Melon diagram approximation}
As discussed in the main text, the FGR type approximation for the Memory matrix (\ref{eq:M-non-interacting}) yields a logarithmically divergent inverse quasiparticle lifetime. Here we fix this divergence with the following non-perturbative approximation for $M$ 
\begin{align}
(M\chi)_{k,\eta,\eta'} & \approx\frac{64\Delta^{2}}{N^{2}}\sum_{p_{1.2}}\cos^{2}(p_{2}/2)e^{ip_{2}(a'-a)}G_{k+p_{1},-\eta,-\eta'}\left[G_{k+p_{1}+p_{2},-\eta,-\eta'}\right]^{*}G_{k+p_{2},\eta,\eta'}\nonumber \\
 & -\frac{16\Delta^{2}}{N^{2}}\sum_{p_{1}p_{2}}(1+e^{ip_{2}})(1+e^{-ip_{1}})\times e^{i(-p_{2})(a)}e^{i(p_{1})a'}[G_{k+p_{1},-\eta,\eta'}][G_{k+p_{1}+p_{2},-\eta,-\eta'}]^{*}[G_{k+p_{2},\eta,-\eta'}]\label{eq:M-melon}.
\end{align}
The difference between (\ref{eq:M-non-interacting}) and (\ref{eq:M-melon}) is that the non-interacting $G$ on the right hand side have been replaced with $G$'s which have been self-consistently calculated by solving (\ref{eqn:Dyson equation real time}). This is represented diagrammatically in the main text. 

To implement this we recast (\ref{eqn:Dyson equation real time}) as a difference equation 
\begin{equation}
    G_{k,\eta\eta'}(t+dt)=G_{k,\eta\eta'}(t)-dt\:\mathrm{i}\epsilon_{\eta\eta''}(k)G_{k,\eta''\eta'}(dt)-dt\int_{0}^{t}d\tau M_{k,\eta\eta''}(\tau)G_{k,\eta''\eta'}(t-\tau).
\label{eq:MMF_k_realtime}
\end{equation}
Assuming we have already found $\{G_{k,\eta\eta'}(s)\}_{s\leq t}$,
we can obtain the solution at the next time-step $G_{k,\eta\eta'}(t+dt)$
by substituting our solutions for earlier times into the RHS, using
the substitution in (\ref{eq:M-melon}) for $M_{k,\eta\eta''}(\tau)$
in terms of the Green's functions at previous time-steps $\{G_{k,\eta\eta'}(s)\}_{s\leq t}$.
In practice we use some tricks to integrate these equations; in particular,
it is numerically more stable to integrate the equations in a rotating
frame, where the second term on the RHS of (\ref{eq:MMF_k_realtime})
disappears. 
\subsubsection{Numerical details}
We solve the difference equation (\ref{eq:MMF_k_realtime}) for
various interaction strengths. In the main text Fig.~\ref{fig:GFDAOE}c shows the results for the $k=0,h=0$ Green's function $G_{k}$, where $k$ is
the wave-vector in the single-site unit cell picture. This is related
to the 2-site unit-cell Green's function through $G_{k}(t)=\frac{1}{2}\sum_{\eta,\eta'}e^{i(\frac{\eta-\eta'}{2})k}G{}_{2k,\eta\eta'}(t)$.  We find the expected logarithmically enhanced decay rate. 

\begin{figure*}
    \centering
    \includegraphics[height=6.5cm]{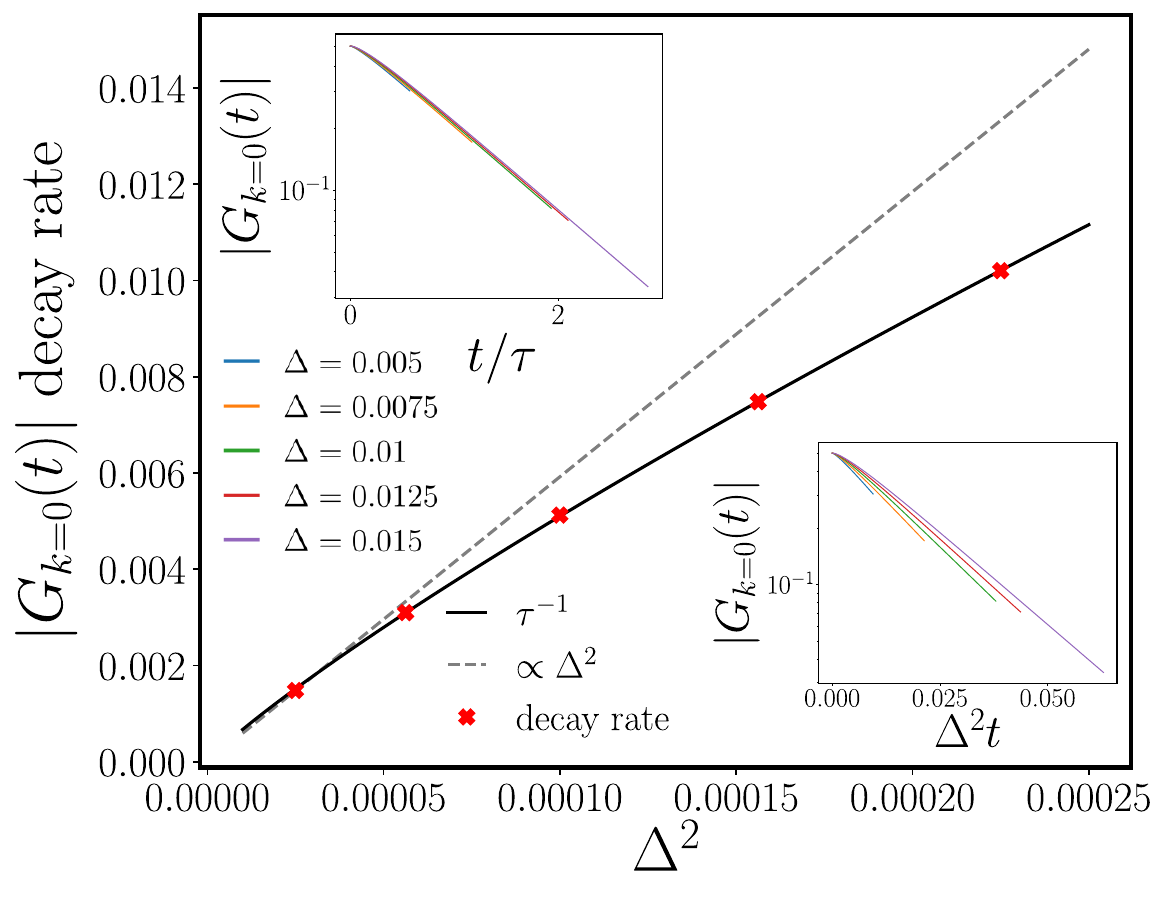}  \includegraphics[height=6.5cm]{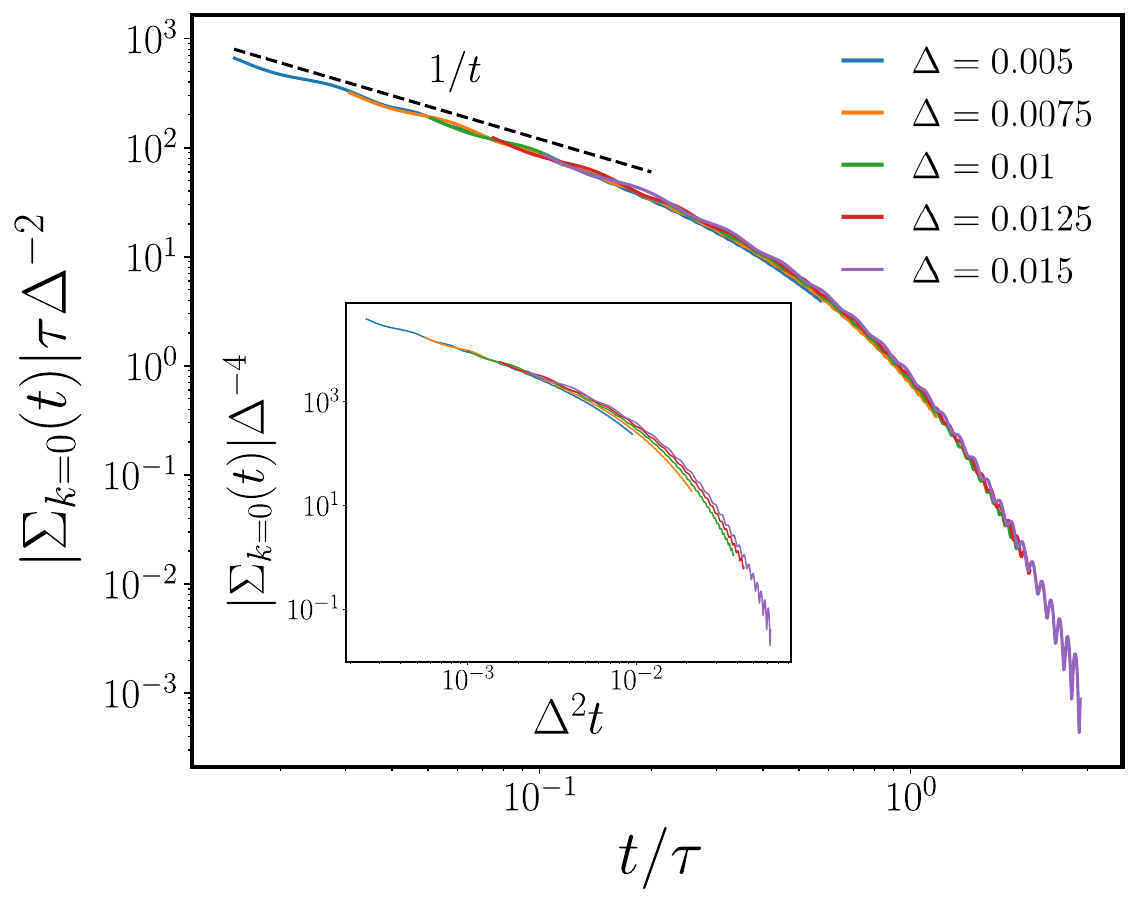}
    \caption{ In this figure, $\tau^{-1} \equiv a\Delta^{2}\log(b\Delta^{-2})$ with $a \approx 6.7, b \approx 0.2$ which is the same as for the fit for Fig.~\ref{fig:GFDAOE} in the main text. (\textit{Left}) Further numerical data  for the decay rate of $G_{k=0}$ against $\Delta^{2}$ for integrable XXZ model with $k=0$ ($L = 800$), using the melonic approximation. Naive Fermi's Golden Rule predicts a straight-line (black dotted). Our modified theory predicts a rate $\propto \tau^{-1}$
(black, solid); this is indeed a better fit to our numerical data
(red X-symbols) obtained by integrating (\ref{eqn:Dyson equation real time}). 
This interpretation is in agreement with the approximate scaling collapse
of $|G_{k=0}(t)|$ vs $t/t_{\Delta}$ in the upper left caption,
and the absence of a scaling collapse of $|G_{k=0}(t)|$ vs $t\Delta^{2}$
in the lower right caption. (\textit{Right}) Single particle self energy for the integrable XXZ model with $k = 0$ ($L=800$). The scaling collapse is consistent with a crossover to superpolynomial decay at a timescale $\mathcal{O}(\tau)$.}
    \label{fig:extra_melon}
\end{figure*}

More generally for $h \neq 0$ we define the `quasiparticle Green's function' via
\begin{equation}
    \tilde{G}_{k,\eta,\eta'}(t) = U_{k,\eta \eta'''}^{\dagger}G_{k.\eta''' \eta''''} (t)U_{k,\eta'''' \eta''}
\end{equation}
where $U_{k}$ is the unitary matrix that diagonalises the matrix $\epsilon_{k}$ describing the free part of the Hamiltonian
\begin{equation}
    \epsilon_{k} = \begin{pmatrix}
2h & -(1/2)(1+e^{\mathrm{-i}k})\\
-(1/2)(1+e^{\mathrm{+i}k}) & -2h
\end{pmatrix}.
\end{equation}
In the absence of interactions, $\tilde{G}_{k}(t) = \mathrm{diag}(e^{i|\omega_{k}|t}/2,e^{-i|\omega_{k}t}/2)$. In the presence of interactions, it will decay. We examine the value of the top-left diagonal element; its decay in time gives a window onto the quasi-particle lifetime of the quasi-particle with energy $\omega_{k} = -\sqrt{(2h)^{2} + \cos^{2}(k/2)}$.
We also perform further tests of the melonic approximation, starting with the integrable XXZ model ($h=0$). First of all, we show that the logarithmically enhanced decay holds for $k=0$ for significantly smaller values of $\Delta$ (Fig.~\ref{fig:extra_melon} (left)). Next, we demonstrate the crossover behavior for the self energy which is predicted to decay as $|\Sigma_{k}(t)|\sim 1/t$ up to a timescale $\mathcal{O}(\tau)$ and then decay superpolynomially at later times. The melonic resummation results are shown in Fig.~\ref{fig:extra_melon} (right) which confirms that the crossover timescale is consistent with the logarithmically enhanced decay rate.

\begin{figure*}
    \centering
    \includegraphics[height=6.5cm]{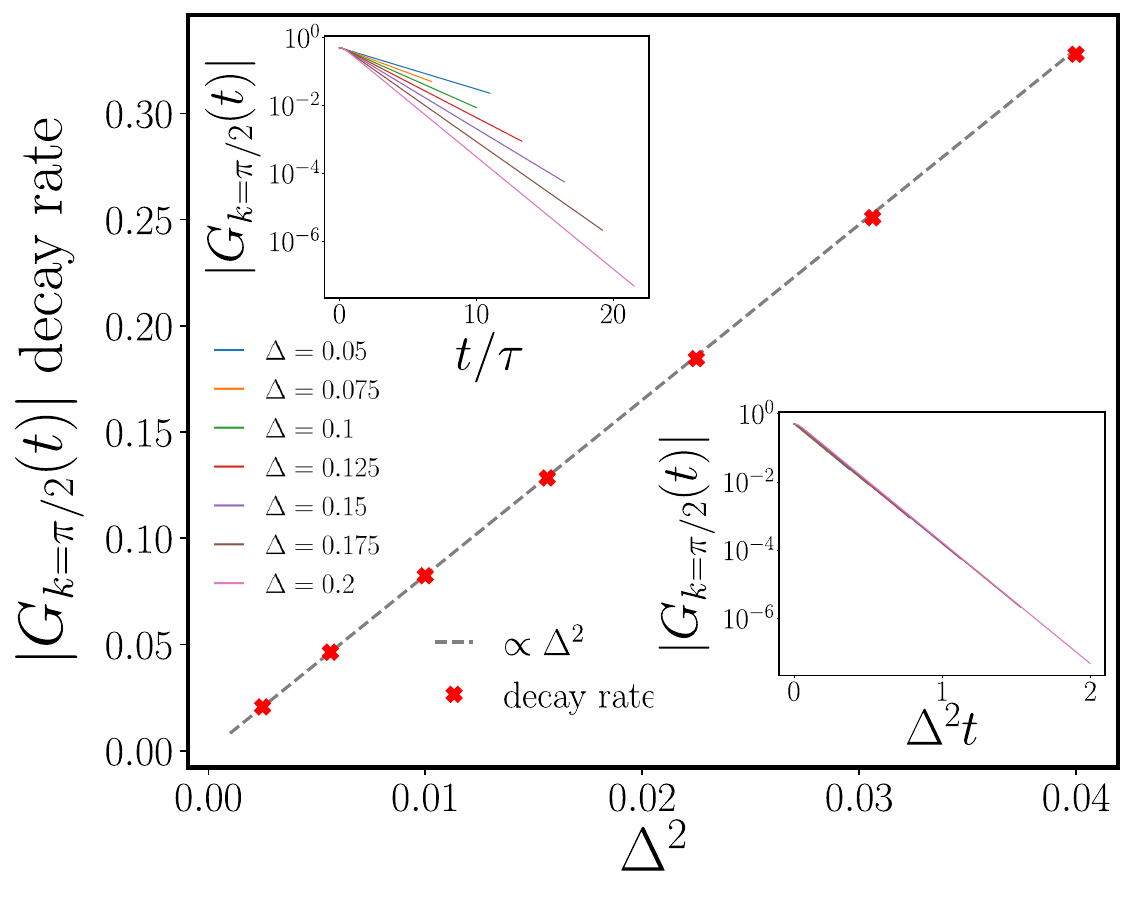} 
    \includegraphics[height=6.5cm]{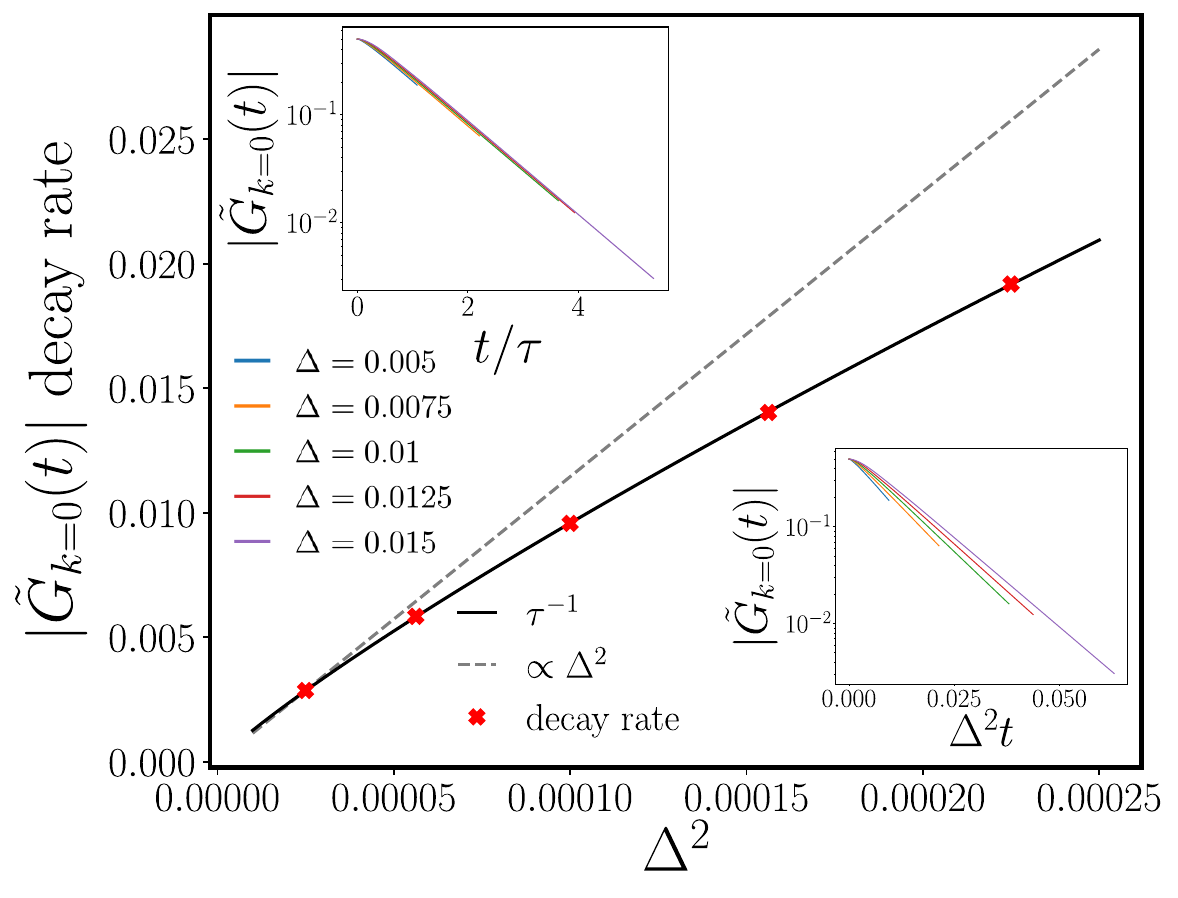} 
    \caption{In this figure, $\tau^{-1} = a\Delta^{2}\log(b\Delta^{-2})$ with $a \approx 6.7, b \approx 0.2$ (left) and $ a \approx 13.1, b \approx 0.15$ (right). (\textit{Left}) Single particle Green's function for the integrable XXZ model with $k = \pi/2$ and $L = 400$. The single particle decay rate is consistent with the non-corrected Fermi's Golden Rule, as expected at this particular wavevector.  (\textit{Right}) single particle Green's function for the non-integrable staggered field XXZ ($h = 0.5$) with $k = 0$ and $L =800$. We find excellent agreement with the logarithmic enhancement to the decay rate that we have previously demonstrated for the integrable XXZ model.} 
    \label{fig:extra_melon2}
\end{figure*}

We also confirm our prediction that the logarithmic enhancement is not present for $k=\pm \pi/2$, with  Fig.~\ref{fig:extra_melon2} (left) showing that the single particle Green's function decays at the usual FGR rate $\Delta^2$. 

Finally, we show that the logarithmically enhanced single particle decay rate also appears in the non-integrable staggered field XXZ model with $h = 0.5$, in Fig.~\ref{fig:extra_melon2} (right). 

The main sources of numerical error are in the Trotter step; for the
numerics displayed here we used $dt=0.1$, and we checked that the
results (e.g., extracted decay rate, exponential decay of Green's
function) did not change qualitatively for a shorter time time simulation
with $dt=0.025$. We also checked for convergence with system size.

\section{Ladder series re-summation}
\begin{figure}[htb!]
    \centering
    \includegraphics[width=0.75\linewidth]{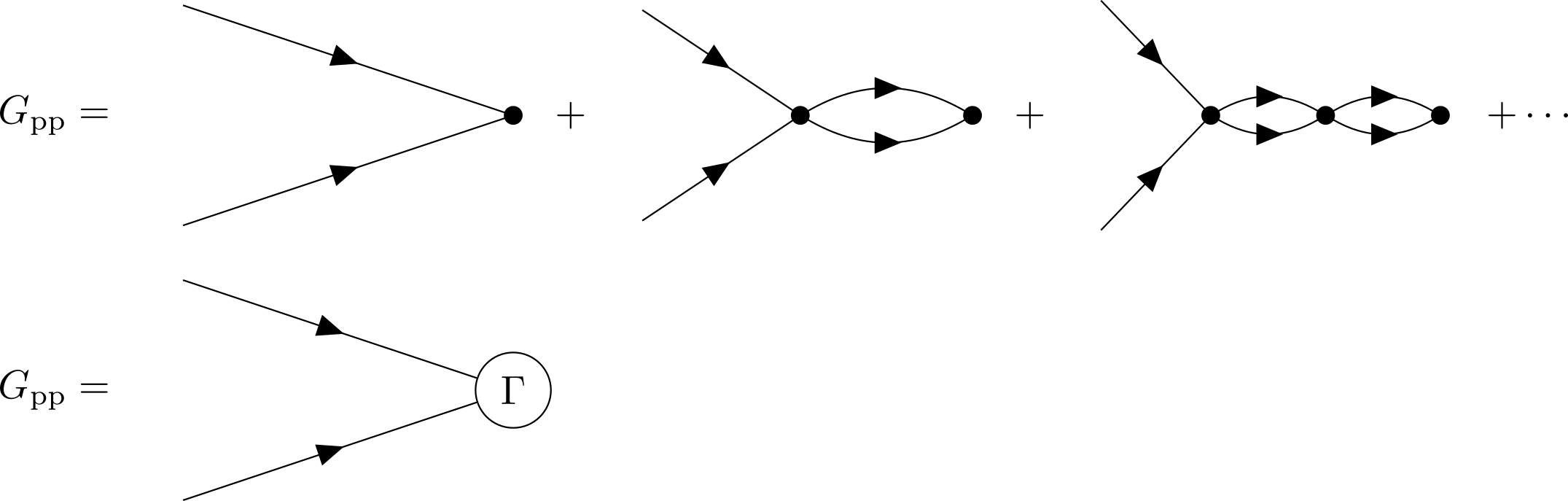}
    \caption{Diagrammatic representation of the 2-body Green's function in terms of the ladder series and the vertex correction to the non-interacting propagator.}
    \label{fig:2-body Green's function}
\end{figure}

In the main text we used the fact that the hole in the ladder series approximation is taken to be non-interacting to express the self-energy in terms of a 2-body propagator 
\begin{equation}
    \Sigma_{k}(z) =  \;2i \int \frac{dqdp}{(2\pi)^{2}}v(q+p-k,q-p-k)G_{\text{pp}}(z,q,p)
\end{equation}
with $v(q,p) = 2i\Delta(\cos q - \cos p)$ being the vertex factor and the `on-shell' behavior being recovered by setting $z = \epsilon_{k} + \epsilon_{2q-k} + i0^{+}$. In the diagrammatic language the 2-body Green's function, $G_{\text{pp}}$ takes the form represented in Fig.~\ref{fig:2-body Green's function}, which can be expressed as a vertex correction to the non-interacting particle-particle propagator $G_{\text{pp}}(z,q,p) = G_{\text{pp}}^{(0)}(z,q,p)\Gamma(z,q,p)$.

\begin{figure}[htb!]
    \centering
    \includegraphics[width=0.5\linewidth]{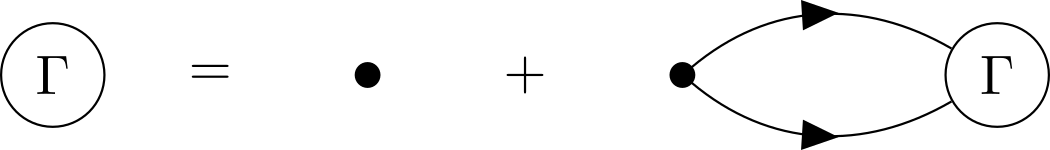}
    \caption{Diagrammatic representation of the vertex correction to the 2-particle green's function. }
    \label{fig:Vertex correction diagram}
\end{figure}

The vertex correction satisfies the following Dyson equation 
\begin{equation}
    \Gamma(z,q,p) = v(q+p-k,q-p-k) + \int \frac{dp'}{2\pi} \; v(p+p',p-p')\Gamma(z,q,p')G_{\text{pp}}^{(0)}(z,q,p')
\end{equation}
represented by Fig.~\ref{fig:Vertex correction diagram}, where the non-interacting particle-particle Green's function is given by 
\begin{equation}
    G_{\text{pp}}^{(0)}(z,q,p) = \frac{1}{4}\frac{i}{z-\epsilon_{q+p} - \epsilon_{q-p}}.
\end{equation}

We can re-cast the Dyson equation as a differential equation to find the self-energy/ quasiparticle lifetime. 
We start by defining the auxiliary function 
\begin{equation}
    P(z,q,x) = \int \frac{dp}{2\pi} \; e^{ipx} G_{\text{pp}}(z,q,p)
\end{equation}
which we can show is simply related to the self-energy
\begin{equation}
\begin{split}
    \Sigma_{k}(z) &= 4\Delta \int \frac{dq}{2\pi} \; \sin(q-k) \int \frac{dp}{2\pi} \; \sin p \;G_{\text{pp}}(z,q,p) \\
    &=  -2i\Delta\int \frac{dq}{2\pi} \; \sin(q-k) \big(P(z,q,1) - P(z,q,-1)\big)
\label{eqn:Self energy P_function}
\end{split}
\end{equation}

In order to solve for the auxiliary function $P$ we first define the differential operator $\hat{L}_{x}$ by the eigenvalue equation 
\begin{equation}
    \hat{L}_{x} e^{ipx} = \frac{1}{G_{\text{pp}}^{(0)}(z,q,p)} e^{ipx}
\end{equation}
which when acted on the auxiliary function yields 
\begin{align*}
    \hat{L}_{x} P(z,q,x) &= \int \frac{dp}{2\pi}\; e^{ipx} \Gamma(z,q,p) \\
    &= \int \frac{dp}{2\pi} \; e^{ipx} v(q+p-k,q-p-k) \; + \int \frac{dpdp'}{(2\pi)^{2}} e^{ipx} v(p+p',p-p')\Gamma(z,q,p')G_{\text{pp}}^{(0)}(z,q,p').
\end{align*}

Remembering that $\Gamma(z,q,p)G_{\text{pp}}^{(0)}(z,q,p) = G_{\text{pp}}(z,q,p)$ allows us to write the second term in terms of $P$
\begin{align*}
     \int \frac{dpdp'}{(2\pi)^{2}} e^{ipx} v(p+p',p-p')\Gamma(z,q,p')G_{\text{pp}}^{(0)}(z,q,p') &= \frac{1}{2}i\Delta \big(\delta(x+1) - \delta(x-1)\big)\big(P(z,q,1) - P(z,q,-1)\big) \\
     &= -\frac{1}{2}i\Delta \big( \delta(x+1) + \delta(x-1)\big)\big(P(z,q,x) - P(z,q,-x)\big)
\end{align*}
where $\delta(x)$ is the Kronecker delta (the final result can also be derived by directly applying the convolution theorem). Using this we obtain the following differential equation for $P$ 
\begin{equation}
    \hat{L}_{x}P(z,q,x) = -\Delta\sin(q-k) \big(\delta(x+1) - \delta(x-1)\big) - \frac{1}{2}i\Delta  \big( \delta(x+1) + \delta(x-1)\big)\big(P(z,q,x) - P(z,q,-x)\big)
\end{equation}

\subsection{Solving the differential equation}
Using the fact that the RHS of the differential equation is odd under $x \to -x$ we find a much simpler differential equation for $P_{-}(z,q,x) \equiv P(z,q,x) - P(z,q,-x)$
\begin{equation}
    \hat{L}_{x}P_{-}(z,q,x) = -2\Delta \sin(q-k) \big(\delta(x+1) - \delta(x-1)\big) - i\Delta  \big( \delta(x+1) + \delta(x-1)\big)P_{-}(z,q,x)
\end{equation}
which is useful since the final quantity we wish to compute (quasiparticle lifetime) is expressed in terms of $P_{-}(z,q,1)$ only. Differential equations of this type can be solved relatively easily in terms of the Green's function for $L_{x}$
\begin{equation}
    \hat{L}_{x} \mathbf{G}(x,x') = \delta(x-x').
\end{equation}
Using the fact that the eigenfunctions of $\hat{L}_{x}$, $e^{ipx}$, form a complete basis and that that the eigenvalues take the form 
\begin{equation}
    \lambda_{p} = \frac{1}{G_{\text{pp}}^{(0)}(z,q,p)}
\end{equation}
we have the following representation for $\mathbf{G}$
\begin{equation}
    \mathbf{G}(x,x') = \int \frac{dp}{2\pi}\;  e^{ip(x-x')} G_{\text{pp}}^{(0)}(z,q,p)  \equiv f(x-x')
\end{equation}

Similarly we define 
\begin{equation}
\begin{split}
    \mathbf{Q}_{0}(x,x') &= \int \frac{dp}{2\pi} e^{ip(x-x')} G_{\text{pp}}^{(0)}(z,q,p) v(q+p-k,q-p-k) \\
    &= -2\Delta \sin(q-k) \big(f(x-x'+1) - f(x-x'-1)\big)
\end{split}
\end{equation}
and define the function $\mathbf{Q}(x,x')$ via 
\begin{equation}
    \hat{L}_{x} \mathbf{Q}(x,x') = \int \frac{dp}{2\pi} e^{ip(x-x')} \; v(q+p-k,q-p-k) - 2i\Delta \big(\delta(x+1) + \delta(x-1)\big)\mathbf{Q}(x,x').
\end{equation}
We notice that the function we wish to compute is related via $P_{-}(z,q,x) = \mathbf{Q}(x,0)$ and that $\mathbf{Q}$ can be simply related to the functions $\mathbf{G},\mathbf{Q}_{0}$ that we can easily compute 
\begin{equation}
    \mathbf{Q}(x,x') = \mathbf{Q}_{0}(x,x') -i\Delta \big( \mathbf{G}(x,1)\mathbf{Q}(1,x') + \mathbf{G}(x,-1)\mathbf{Q}(-1,x')\big). 
\end{equation}

Setting $x = \pm 1$ yields a set of 2 coupled equations for $\mathbf{Q}(\pm1,x')$ that can be solved to give 
\begin{equation}
    P_{-}(z,q,1) = \mathbf{Q}(1,0) = \frac{2\Delta \sin(q-k)(f(0)-f(2))}{1 + i\Delta(f(0)-f(2))}
\label{eqn:P_explicit}
\end{equation}
in doing so we have used the fact that for integer $n$, $f(-n) = f(n)$.

We recall the definition of the function $f(n)$ from which the quasiparticle lifetime is computed 
\begin{equation}
    f(n) = \frac{i}{4} \int \frac{dp}{2\pi} \frac{e^{ipn}}{z - \epsilon_{q+p} - \epsilon_{q-p}}
\end{equation}
with $\epsilon_{k} = -\cos k$.
Performing the change of variables $w = e^{ip}$ yields a contour integral around the unit circle 
\begin{equation}
    f(n) = \frac{1}{4\cos q} \oint \frac{dw}{2\pi} \frac{w^{n}}{w^{2} + wz/\cos q + 1}.
\end{equation}
Taking the 'on-shell' limit to compute the quasiparticle lifetime amounts to setting $z = \epsilon_{k} + \epsilon_{2q-k} + i0^{+}$. Performing the contour integral in this limit via the residue theorem (taking special care to pick the correct root depending on the sign of $\cos q$) yields 
\begin{equation}
\begin{split}
    f(0) &= \frac{1}{8|\cos q\sin(q-k)|} \\
    f(2)&= \frac{\cos^{2}(q-k) - \sin^{2}(q-k)}{8|\cos q \sin(q-k)|} - i\frac{\cos(q-k)}{4\cos q}.
\end{split}
\end{equation}
Plugging these results into the integral expression for the self-energy (\ref{eqn:P_explicit}, \ref{eqn:Self energy P_function}) yields 
\begin{equation}
    \tau_{k}^{-1} = 16\Delta^{2} \int \frac{dq}{2\pi} \frac{|\sin^{3}(q-k)||\cos q|}{\left(4\cos q- \Delta \cos(q-k)\right)^{2} + \left(\Delta \sin(q-k)\right)^{2}}
\end{equation}
from which we obtain the expression in the main text for $\Delta \to 0$.

\section{Classical Floquet model}
\begin{figure}[htbp!]
    \centering
    \includegraphics[width=0.5\linewidth]{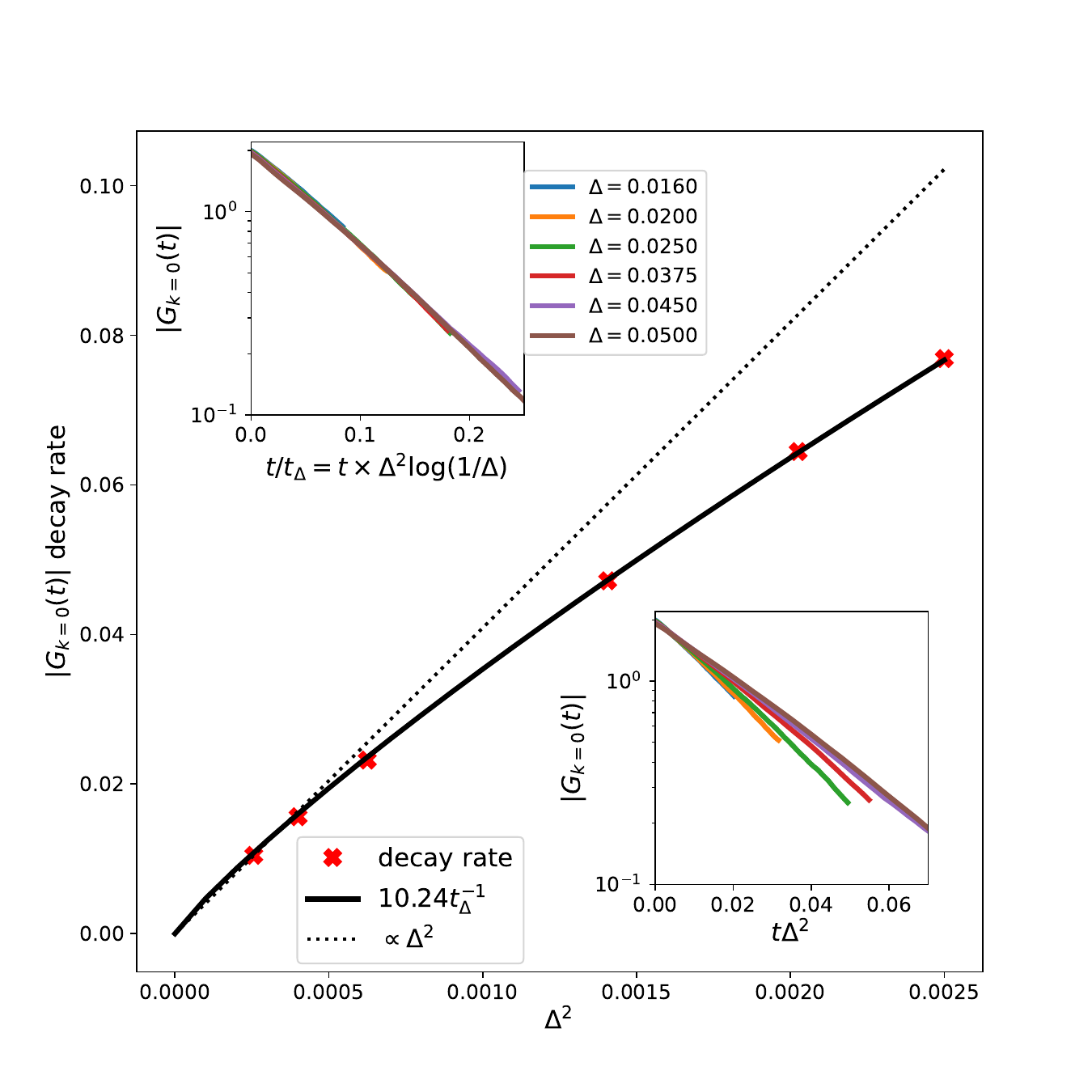}
    \caption{Summary of autocorrelator ($k=0$) decay rates as a function of $\Delta$, for the classical Floquet model. Insets  are consistent with exponential decay. The data for $\Delta\protect\geq0.0375$ used $120$K sample averages and converged with system size for $L=361$. The smaller $\Delta$ appeared to converged for $20$K samples, and only required $L=181$.}
    \label{fig:Classical summary}
\end{figure}
Consider a classical many-body problem with phase space $\mathbb{C}^{L}$, which is isomorphic to the phase space for $L$ coupled 1d oscillators $(p_{x},q_{x})_{x = 1}^{L}$. To draw comparisons with the quantum models that we consider we work in terms of the complex fields $\psi_{x} = (q_{x}-ip_{x})/\sqrt{2}$, which has the classical Poisson bracket
\begin{equation}
    \left\{\psi_{x}^{*},\psi_{y}\right\} = i\delta_{xy}. 
\end{equation}
The dynamics are defined by a two step Floquet evolution respecting $U(1)$ symmetry $\psi \to e^{i\theta}\psi$. The first step of the evolution is 

\begin{equation}
\begin{split}
\psi_{x}(t+1/2) & =(e^{i\{H_{0},\cdot\}}\psi_{x})(t)\\
 & =(e^{i\boldsymbol{h}_{0}}\psi)_{x}(t),
\end{split}
\end{equation}
where 
\begin{equation}
    H_{0} = \sum_{rr'}(h_{0})_{rr'} \psi_{r}^{*}\psi_{r'}
\end{equation}
is a real quadratic classical Hamiltonian. The second step of the evolution 
\begin{equation}
\begin{split}
    \psi_{x}(t+1) &= (e^{i\left\{ \Delta\sum_{r}|\psi_{r}|^{4},\cdot \right\}}\psi_{x})(t+1/2) \\
    &= (e^{i\Delta\boldsymbol{h}_{\text{I}}(\psi)}\psi)_{x}(t),
\end{split}
\end{equation}
where 
\begin{equation}
    (h_{\text{I}}(\psi))_{rr'} = 2|\psi_{r}|^{2}\delta_{rr'}
\end{equation}
corresponds to evolution with the real quartic classical Hamiltonian 
\begin{equation}
    H_{\text{I}} = \Delta \sum_{r} |\psi_{r}|^{4}.
\end{equation}
In our numerical simulation Fig.~\ref{fig:Classical summary} we calculate correlators of the form $C_{x}(t) = \langle \psi_{x}(t)\psi_{0}^{*}(0)\rangle$, sampling initial states from the Gaussian ensemble 
\begin{equation}
    \rho \propto \; e^{-\sum_{r}|\psi_{r}|^{2}/2}.
\end{equation}
The correlation function $C_{k}(t) \propto \sum_{x} e^{-ikx}C_{x}(t)$ is the classical analogue of the single particle Green's function considered in the main text, involving the autocorrelator of a charged local operator. We find that to the times we are able to simulate, $C_{k}(t)$ decays super-polynomially in time and moreover that the decay rate is much better fit by $\Delta^{2} \log \Delta^{-2}$ than by $\Delta^{2}$, much as we find the for the quantum mechanical systems studied in the main text. 

\section{DAOE method}
\begin{figure}
	
\includegraphics[height=5.5cm,keepaspectratio]{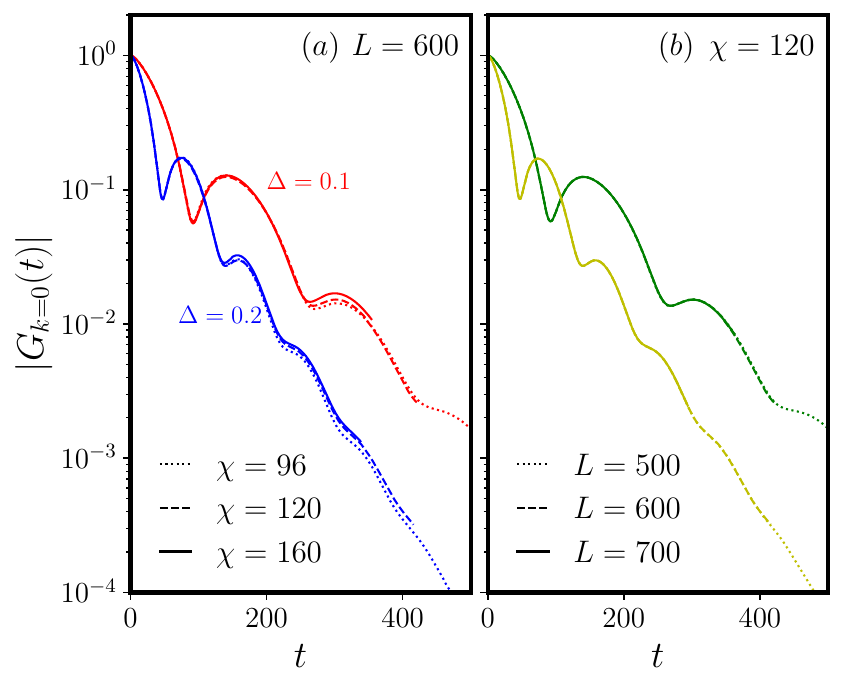}%
  \includegraphics[height=5.5cm,keepaspectratio]{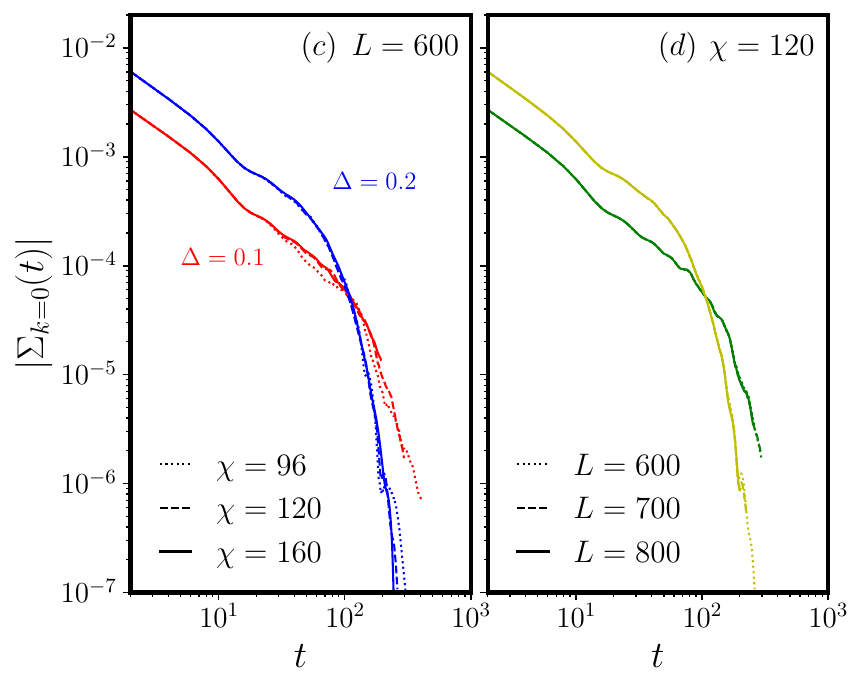}%
  \includegraphics[height=5.5cm,keepaspectratio]{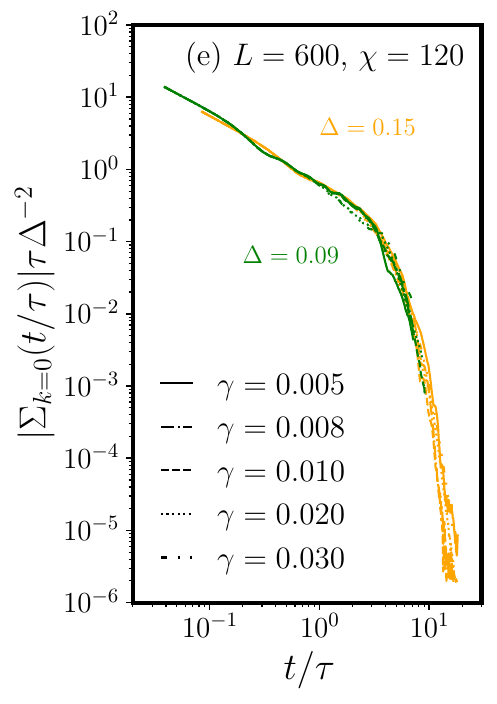}
\caption{(a) Convergence of Green's function data with bond dimension $\chi$, for fixed $L=600$. (b) Convergence of Green's function data with system size $L$, for fixed $\chi=120$. (c) Convergence of self-energy data with bond dimension $\chi$, for fixed $L=600$. (d) Convergence of self-energy data with system size $L$, for fixed $\chi=120$. (e) Self-energy collapse for different values of $\gamma$.\label{fig:bdLconvergence}} 
\end{figure}

In the DAOE method \cite{DAOE1}, an operator is evolved with a standard Time Evolving Block Decimation (TEBD) algorithm \cite{vidal2003efficient}, but with an additional ``dissipative'' step used to periodically truncate components of the operator with large spatial support. The dissipation effectively multiplies each Pauli-string within the evolved operator by a factor $\propto \exp(-\gamma \text{max}(0,\ l-l^*))$, where $l$ is a measure of the operator size, and $l^*$ is a cutoff length-scale under which operators are not truncated. The unitary limit can be recovered by extrapolating in $\gamma \to 0$ or $l^* \to \infty$. By suppressing operator components with large spatial support, the evolution is effectively restricted to a subspace of (polynomially many) operators, avoiding the entanglement barrier of tracking evolution in the entire Hilbert space. The DAOE algorithm was extended to fermionic problems in \cite{jerome_daoe}. Application to odd-parity fermion strings used in this work requires simply starting in the ``$1^-$" state of the DAOE MPO, see Fig. 2 in Reference \cite{jerome_daoe}, and requires an MPO bond dimension of $D=l^*+2$. 

For the Green's function calculation, we represent $G_k(t) = \langle f_k(t) | f^\dagger_k(0) \rangle$ in terms of the `vectorised' fermion operators $|f_k(t)\rangle = |U^\dagger(t) f_k U(t)\rangle$, with the operator inner product $\langle A|B\rangle = \text{Tr}(A B)/\mathcal{N}$, and $\mathcal{N}$ the Hilbert space dimension (we work at infinite temperature throughout). In the simulations in the main text, we fix the system size $L=600$, use a second-order TEBD time step $\delta t = 0.1$, maximum MPS bond dimension $\chi=120$, and DAOE parameters $l^*=3$ and a DAOE dissipation period $\delta t_\gamma = 1$. We also fix the prefactor of the Hamiltonian hopping term in Eq.~\eqref{eqn:Hamiltonian} to $J=0.2$: all values of $\Delta$ reported in the text are relative to $J$. In Fig.~\ref{fig:GFDAOE}a we plot the resulting Green's function against the rescaled time $t/\tau$, with an extrapolation of the data in the DAOE dissipation rate $\gamma \to 0$, using a linear fit and the points $\gamma = 0.005, 0.008, 0.01, 0.02, 0.03$. To ascertain convergence in the parameters, in Fig.~\ref{fig:bdLconvergence}a-b we show the convergence in bond dimension for fixed system size, and in system size for fixed bond dimension, for two values of $\Delta$ used in the main text.

For the self-energy, an exact expression is given by Eq.~\eqref{eq:self-energy-mem}. This is an autocorrelation function restricted to the space of fast operators i.e., operators which contain more than one fermion operator. Up to unimportant multiplicative constants, this is
\begin{equation}
     \mathrm{i}\Sigma_k(t)  = \langle f_{k}^{\dagger}|\mathcal{L}Qe^{-iQ\mathcal{L}Qt}Q\mathcal{L}|f_{k}^{\dagger}\rangle.
\end{equation}
It is numerically difficult to implement the time evolution $e^{-iQ\mathcal{L}Qt}$ which corresponds to constantly strongly projecting the dynamics into the fast space. Instead, we consider a discretised version of this quantity which we expect to behave similarly; we periodically (rather than constantly) project the dynamics into the fast space
\begin{equation} \label{eq:poormanselfenergy} \mathrm{i}\Sigma_k(t) \approx  \lim_{\delta t\to 0}\bra{f^\dagger_k}{\mathcal{U}(\mathcal{Q}\mathcal{U}\mathcal{Q}})^{t/\delta t} \mathcal{U}\ket{f^\dagger_k}.
\end{equation}
This expression can then be treated numerically, however the uncontrolled time-evolution poses the same issues as vanilla TEBD. We thus use the ``dissipated'' version of the self-energy shown in Eq.~\eqref{eq:SEDAOE}, by accompanying each appearance of $\mathcal{Q}$ with a soft DAOE dissipation factor. We use the same parameters as for the Green's function simulations, but fix the dissipation rate $\gamma = 0.01$. We checked that other values produced essentially the same results (see Fig.~\ref{fig:bdLconvergence}e), though a clean extrapolation in $\gamma \to 0$ was not possible for our data. Nevertheless, the self-energy in Fig.~\ref{fig:GFDAOE}b displays the predicted features: a $1/t$ decay up to times of order the log-enhanced lifetime $\tau$, followed by a superpolynomial decay at later times. This further holds in the non-integrable $h\neq0$ regime, see Fig.~\ref{fig:SENI}, below. In Fig.~\ref{fig:bdLconvergence}c-d we show convergence data for the self-energy, as for the Greens function data.

\section{Additional data from DAOE simulations}

\begin{figure}
    \includegraphics[width=0.5\linewidth]{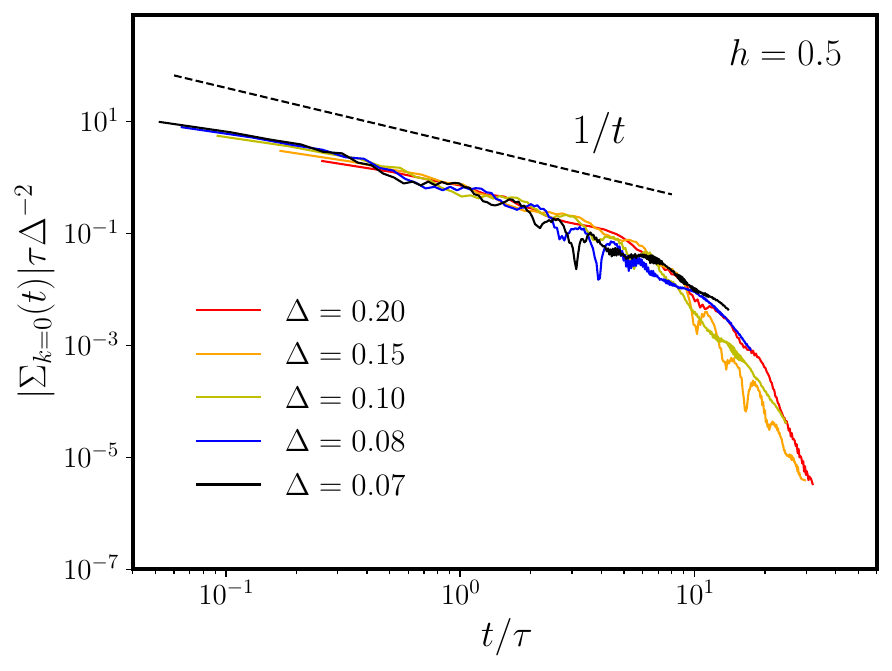}
    \caption{Single-particle self-energy $\Sigma_k(t)$ vs.~the rescaled time, for the non-integrable model with $h=0.5$. Here the DAOE parameter $\gamma=0.01$. The data is consistent with the correction to Fermi's Golden Rule, as shown in the main text for the integrable model.}
    \label{fig:SENI}
\end{figure}

\begin{figure}
    \includegraphics[width=0.55\linewidth]{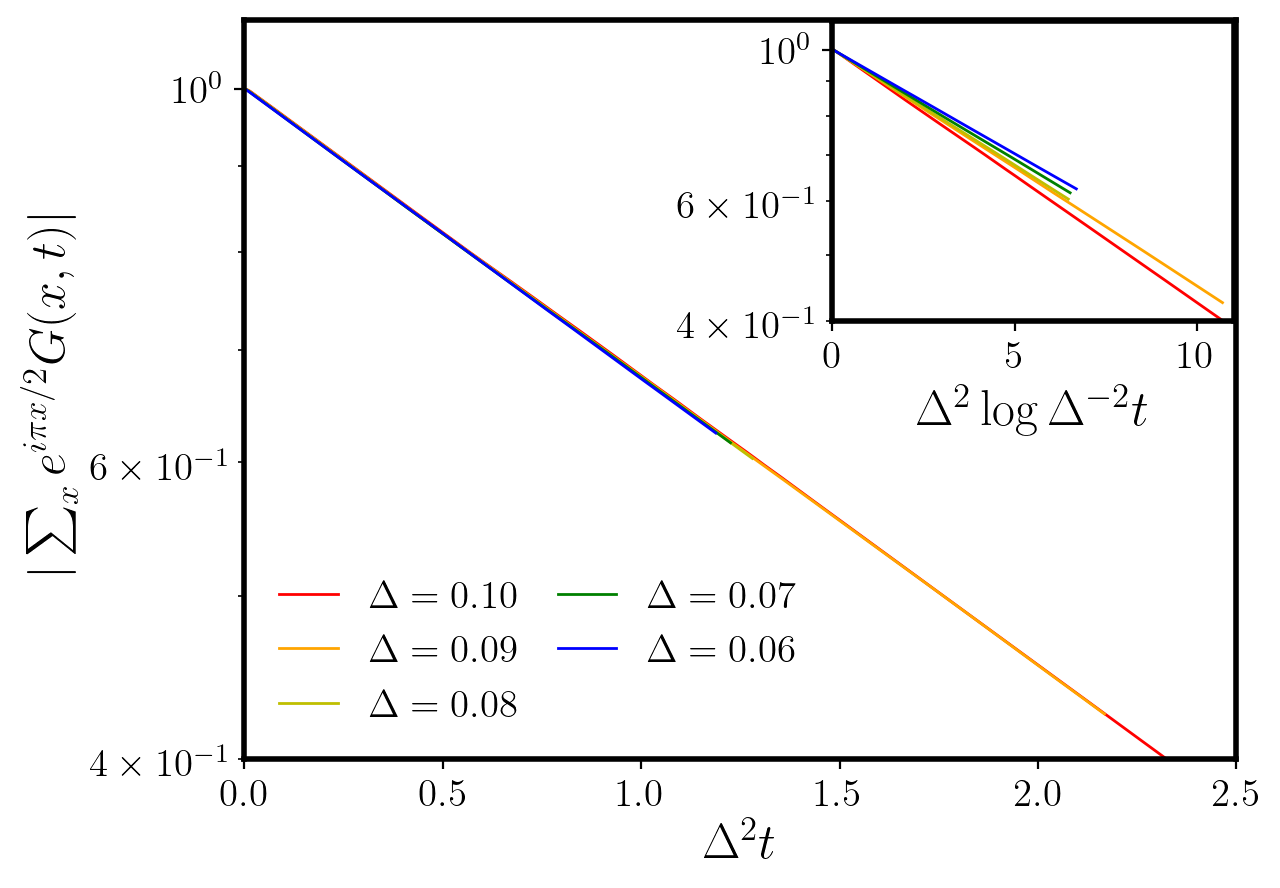}
    \caption{Decay of real-space Greens function summed with oscillating phase $e^{i\pi x/2}$, which is approximately equal to $|G_{k=\pi/2}(t)|$. The function decays according to the non-corrected Fermi's golden rule prediction, as expected. \emph{Inset}: same data scaled according to $\tau^{-1} = \Delta^2 \log \Delta^{-2}$.}
    \label{fig:pi_decay}
\end{figure}
First we show that the behaviour of the self-energy predicted by Eq.~\eqref{eqn:Lifetime self consistent equation} holds analogously in a non-integrable model. We consider the same Hamiltonian as used in the main text, Eq.~\eqref{eqn:Hamiltonian}, with an additional staggered field $h$ which breaks the model's integrability:

\begin{equation}
    H = -\frac{1}{2}\sum_{i} \left(f_{i+1}^{\dagger}f^{\null}_{i} + f_{i}^{\dagger}f^{\null}_{i+1}\right) + \Delta \sum_{i} n_{i+1}n_{i} - h\sum_i (-1)^in_i.
\label{eqn:Hamiltonian_NI}
\end{equation}
We show the results for $k=0$ and with a field $h=0.5$, in Fig.~\ref{fig:SENI}. Other parameters are the same as in Fig.~\ref{fig:GFDAOE} for the self-energy. We confirm that the self-energy decays according to $1/t$ up until times of order $\tau$, followed by a superpolynomial decay. This is consistent with the arguments given in the main text, which do not rely on the integrability of the model in any way.

In the main text, it was also noted that the logarithmic correction to the lifetime is not expected at the exceptional point $k=\pi/2$. This was also confirmed in the re-summed solution in the ladder summation, see Eq.~\eqref{eqn:Quasiparticle lifetime ladder series}. We checked that the real-space Greens function $G(x,t)$, when summed with the oscillating phase $e^{i\pi x/2}$, decays according to the Fermi's golden rule prediction, i.e.~ $|\sum_x e^{i\pi x/2} G(x,t)| \propto e^{-\Delta^2 t}$. Note that since we work with open boundary conditions this function is not exactly equal to the $G_{k=\frac{\pi}{2}}(t)$ Green's function; however the difference is negligible for large system sizes. The data is shown in Fig.~\ref{fig:pi_decay}, while the inset shows that the scaling $\tau^{-1} = \Delta^2 \log \Delta^{-2}$ provides the wrong result for the point $k=\pi/2$. Our numerics uses a minimum system size of $L=400$.


\end{document}